\documentclass[11pt,one column]{IEEEtran} 

\usepackage{graphicx}
\usepackage{subfig, caption}
\usepackage{color}
\usepackage{epsfig, multicol}
\usepackage{amssymb,amsmath,amsthm}
\usepackage{latexsym}
\usepackage{mathrsfs}
\usepackage{bbm, ulem}
\usepackage{setspace} 
\usepackage{epstopdf, titling}
\usepackage{mathtools}
\usepackage[hang,flushmargin]{footmisc}
\usepackage{flushend}
\usepackage{fancybox}
\usepackage{relsize}
\usepackage{lmodern}
\mathtoolsset{showonlyrefs}
\newcommand{\cJ}{{\mathcal J}}
\newcommand{\cX}{{\mathcal X}}
\newcommand{\cY}{{\mathcal Y}}

\newcommand{\cM}{{\mathcal M}}

\newcommand{\cAk}{{\mathcal A}_{k}}

\newcommand{\cXm}{{\mathcal X}_{\mathcal M}}
\newcommand{\cYm}{{\mathcal Y}_{\mathcal M}}
\newcommand{\cU}{{\mathcal U}}

\newcommand{\ep}{\epsilon}

\newcommand\MC{{ \ - \!\!\circ\!\! - \ }}

\newtheorem{theorem}{Theorem}
\newtheorem{proposition}[theorem]{Proposition}
\newtheorem{corollary}{Corollary}

\newtheorem{lemma}[theorem]{Lemma}

\theoremstyle{remark}
\newtheorem*{remark*}{Remark}
\newtheorem*{remarks*}{Remarks}

\theoremstyle{definition}
\newtheorem{example}{Example}
\newtheorem{definition}{Definition}

\date{}

\allowdisplaybreaks
\begin{document}

\title{ {\bf Sampling Rate Distortion} }
\author{Vinay Praneeth Boda and Prakash Narayan$^\dag$   }
\maketitle {\renewcommand{\thefootnote}{}\footnotetext{
\vspace{.02in}

\noindent 
$^\dag$V.P. Boda and P. Narayan are with the Department of
Electrical and Computer Engineering and the Institute for Systems
Research, University of Maryland, College Park, MD 20742, USA.
E-mail: \{praneeth, prakash\}@umd.edu. 

This work has been presented
in parts at the IEEE International Symposia on Information Theory in
2014 \cite{BodaNarayan14} and 2016 \cite{BodaNarayan16}; and at the $53^{rd}$
Annual Allerton Conference on Communication, Control and Computing in 
2015 \cite{BodaNarayan15}.

\noindent It was supported by the U.S.
National Science Foundation under Grants CCF-0917057 and CCF-1319799.}}

\thispagestyle{plain}
\pagestyle{plain}

\vspace*{-0.5 cm}
\begin{abstract}

Consider a discrete memoryless multiple source with $m$ components of which  $k \leq m$ possibly different sources are sampled at each time instant and jointly 
compressed in order to reconstruct all the $m$ sources under a given distortion criterion.
A new notion of sampling rate distortion function is introduced, and is characterized first for the case of fixed-set sampling. Next, for independent random sampling performed without knowledge of the source outputs, it is shown that the sampling rate distortion
function is the same regardless of whether or not the decoder is informed of the sequence of sampled sets. Furthermore,  memoryless random sampling is considered with the sampler depending on the source outputs and with an informed decoder. It is shown that deterministic sampling, characterized by a conditional point-mass, 
is optimal and suffices to achieve the sampling rate distortion function. For memoryless random sampling with an
uninformed decoder, an upper bound for the sampling rate distortion function is seen to possess a similar property of conditional point-mass optimality. It is shown by example that memoryless sampling with an informed decoder can outperform strictly any independent random sampler, and that memoryless sampling can do strictly better with an informed decoder than without.
\end{abstract}


\section{Introduction}

Consider a set $\cM$ of $m$ discrete memoryless sources with a known joint probability mass function. Subsets of $k \leq m$ sources are sampled at each time instant, and jointly processed with the objective of 
reconstructing {\it all} the $m$ sources as  compressed representations, within a specified level of distortion. How should the sampler optimally sample the sources in a causal manner to yield the best compression rate for a given distortion level? What are the tradeoffs -- under optimal processing -- among the sampling procedure, compression rate and distortion level? This paper is our preliminary attempt at answering these questions.

\vspace*{0.1cm}

The study of problems of combined sampling and compression has a rich and varied history in diverse contexts. Highlights include: classical sampling and processing, rate distortion theory, multiterminal source coding, wavelet-based compression, and compressed sensing, among others. Rate distortion theory \cite{Ber71} rules the compression of a given sampled signal and its reconstruction within a specified distortion level. On the other hand, compressed sensing \cite{Don06,CanRomTao06,CanTao06} provides a random linear encoding of nonprobabilistic analog sources marked by a sparse support, with lossless recovery as measured by a block error probability (with respect to the distribution of the
encoder). Upon placing the problem of lossless source coding of analog sources in an information theoretic setting,
with a probabilistic model for the source that need not be encoded linearly, R\'enyi dimension is known to determine fundamental performance limits \cite{WuVer10} (see also \cite{KawDem94, WuVer12}). Several recent
studies consider the compressed sensing of a signal with an allowed detection error rate or quantization
distortion \cite{FleRanGoy07, ReeGas12, WeiVet12}, or with denoising \cite{FleRanGoyRam06}; of multiple signals followed by distributed quantization \cite{SunGoy12}, including a study of scaling laws \cite{IshKunRam03}; or of sub-Nyquist rate sampled
signals followed by lossy reconstruction \cite{KipnisGold16}. 
\vspace*{0.1cm}

Closer to the line of our work, the rate distortion function has been characterized when multiple Gaussian signals from a random field are  sampled and quantized (centralized or distributed) in \cite{PraNeu07, NeuPra11, NeuPra12}. Also, in a series \cite{UnnVet12a, UnnVet12b, UnnVet12c} (see also \cite{HorRiyLuVet10}, \cite{RanVet13, ChenRanZhanVet13}), various aspects of random field-sampling and reconstruction for special models are considered. In a setting of distributed acoustic sensing and reconstruction, centralized as well as distributed coding schemes and sampling lattices are studied, and their performance is compared with corresponding rate distortion bounds  \cite{KonTelVet12}. In \cite{KashLasXiaLiu05}, considering a Gaussian random field on the interval $[0,1]$ and i.i.d. in time, reconstruction of the entire field from compressed versions of $k$ sampled sequences under the mean-squared error distortion criterion is studied. In a different formulation, for the case of $m=2$ sources, each of which is sampled for a fixed proportion of time, the rate distortion function and associated sampling mechanism are characterized in \cite{LiuSimErk12}.

\vspace*{0.1cm}

Our work differs from the approaches above in that we allow randomized sampling that can depend on the observed source values, and no sparsity assumption is made on the sources. It bears emphasis that we deal with centralized --  and not distributed -- processing of the sources.

\vspace*{0.1cm}

Our contributions are as follows. We consider a new formulation involving a {\it sampling rate distortion\footnote{This apt terminology has been used also in an earlier work on compressed sensing with error tolerance \cite{ReeGas12}.} function} (SRDf), which combines a sampling of sources and lossy compression, to address the questions posed at the outset. 
As a basic ingredient, the sampling rate distortion function is characterized for a fixed sampling set of size $k \leq m$. This characterization is based on prior work by Dobrushin-Tsybakov \cite{DobTsy62} (see also Berger \cite{Ber71}, \cite{Ber78} and Yamamoto-Itoh \cite{YamIto80}) on the rate distortion function for a ``remote'' source-receiver model in which the encoder and receiver lack direct access to the source and decoder outputs, respectively. For the special case of the probability of error distortion criterion, we show that the optimal procedure can be simplified to a rate distortion code for the sampled sources followed by maximum a posteriori estimation of the remaining sources. 

\vspace*{0.1cm}

Best fixed-set sampling can be strictly inferior to random sampling.  Considering an independent random sampler, in which the sampling does not depend on the source outputs and is independent  (but not necessarily identically distributed) in time, we show that the corresponding  SRDf remains the same regardless of whether or not the decoder is provided information regarding the sequence of sampled sets. This surprising property does not hold for any causal sampler, in general.
Next, we consider a generalization, namely a memoryless random sampler whose output can depend on the source values at each time instant. The associated formula for SRDf  is used now to study the structural characteristics of the optimal sampler. Specifically, we show when the decoder too is aware of the sequence of sampled sets that the optimal sampler is characterized by a conditional point-mass; this has the obvious benefit of a reduction in the search space for an optimal sampler. We also show that such a memoryless sampler can outperform strictly a random sampler that lacks access to source values. Finally, in a setting in which the decoder is unaware of the sampled sequence, an upper bound for the SRDf is seen to have an optimal conditional point-mass sampler.
 
\vspace*{0.1cm}

Our models are described in Section \ref{s:Preliminaries}. The main results, along with examples, are stated in Section \ref{s:Results}.  Section \ref{s:Proofs} contains the proofs. Presented first are the achievability proofs that are built successively in the order of increasing complexity of the samplers. The converse proofs follow in reverse order in a unified manner.
 

\section{Preliminaries}\label{s:Preliminaries}

Let $\cM = \{ 1, \ldots, m \}$ and $X_{\cM} = ( X_{1}, \ldots, X_{m} )$ be a $\cX_{\cM} = \mathop{\mbox{\large $\times$}} \limits_{i=1}^m \cX_{i}$-valued rv where each $\cX_{i}$ is a finite set. It will be convenient to use the following compact notation. For a nonempty set $A \subseteq \cM $, we denote by $X_{A}$ the rv $ ( X_{i}, i \in A ) $ with values in $\mathop{\mbox{\large $\times$}} \limits_{i \in A} \cX_{i}$, and denote $n$  repetitions of $X_{A}$ by $X_{A}^{n} = ( X_{i}^{n}, i \in A)$ with values in $\cX_{A}^{n} = \mathop{\mbox{\large $\times$}} \limits_{i \in A} \cX_{i}^{n}$, where $X_{i}^{n} = ( X_{i1}, \ldots, X_{in})$ takes values in the $n$-fold product space ${\cal X}_{i}^{n} = \cX_{i} \times \cdots \times \cX_{i}$. For $1 \leq k \leq m$, let ${\cal A}_{k} = \{ A:  A \subseteq \cM, \ |A| = k \}$ be the set of all  $k$-sized subsets of $\cM$ 
and let $A^{c}  = \cM \setminus A$. All logarithms and exponentiations are with respect to the base 2.

\vspace*{0.1cm}

Consider a discrete memoryless multiple source (DMMS) $\{ X_{\cM t} \}_{t=1}^{\infty}$  consisting of i.i.d. repetitions of the rv $X_{\cM}$ with given pmf $P_{X_{\cM}}$ of assumed full support ${\cal X}_{\cM}$. Let $\cY_{\cM} = \mathop{\mbox{\large $\times$}} \limits_{i=1}^m \cY_{i},$ where $\cY_{i}$ is a finite reproduction alphabet for $X_{i}$.

\begin{definition}\label{d: RS}
 A $k$-{\it random sampler} ($k$-RS), $1 \leq k \leq m$, collects causally at each $t=1, \ldots, n$, random samples$ ^\dag${\renewcommand{\thefootnote}{}\footnotetext{ $^\dag$With an abuse of notation, we write $X_{S_{t} t}$ simply as $X_{S_t}.$}} $X_{S_{t}}$ from $X_{\cM t}$, where $S_{t}$ is a rv with values in $\cAk$ with (conditional) pmf $P_{S_{t} | X_{\cM}^{t} S^{t-1}} $, with $X_{\cM}^{t} = (X_{\cM 1}, \ldots, X_{\cM t} )$ and $S^{t-1} = ( S_{1}, \ldots , S_{t-1} ) $. Such a $k$-RS is specified by a (conditional) pmf $P_{S^{n}|X_{\cM}^{n}}$ with the requirement \begin{equation}
  \label{eq:k-RS-distribution}
 P_{S^{n} | X_{\cM}^{n}} = \prod_{t=1}^{n} P_{S_{t} | X_{\cM}^{t} S^{t-1} }.                                                                                                                                                                                                                                                                                                                                                                                                                                                                                                                                                                                                                                                                                                                       \end{equation}
The output of a $k$-RS is $(S^{n},X_{S }^{n})$ where $X_{S }^{n} = (X_{S_{1} },\ldots,X_{S_{n} })$. Successively restrictive choices of a $k$-RS in \eqref{eq:k-RS-distribution} corresponding to 
 \begin{equation}
 \label{eq:k_MRS_def}
  P_{S_{t}|X_{\cM}^{t} S^{t-1}} = P_{S_{t}|X_{\cM t}}, \ \ t=1, \ldots,n
 \end{equation}
and 
\begin{equation}
\label{eq:k_IRS_def}
 P_{S_{t}|X_{\cM}^{t} S^{t-1}} = P_{S_{t}}, \ \ \ \ \ \  t=1, \ldots,n
\end{equation}
will be termed the $k$-{\it memoryless} and the $k$-{\it independent random samplers} and denoted by $k$-MRS and $k$-IRS, respectively.
 
\end{definition}


\begin{definition} \label{d:encoder}
 An $n$-length block code with $k$-RS for a DMMS $\{ X_{\cM t} \}_{t=1}^{ \infty} $ with alphabet $\cX_{\cM}$ and reproduction alphabet $\cY_{\cM}$ is a triple $(P_{S^{n}|X_{\cM}^{n}}, f_{n}, \varphi_{n} )$ where $P_{S^{n} | X_{\cM}^{n} }$ is a $k$-RS as in \eqref{eq:k-RS-distribution}, and  $(f_{n}, \varphi_{n} )$ are a pair of mappings where the encoder $f_{n}$ maps the output of the $k$-RS into some finite set $\cJ = \{1, \ldots , J \}$ and the decoder $\varphi_{n}$ maps $\cJ$ into $\cY_{\cM}^{n}$. We shall use the compact notation $(P_{S|X_{\cM}}, f, \varphi),$ suppressing $n$. The rate of the code with $k$-RS $(P_{S|X_{\cM}}, f, \varphi)$ is $\dfrac{1}{n} \log J$.
\end{definition}


For a given (single-letter) finite-valued distortion measure $d: \cX_{\cM} \times \cY_{\cM} \rightarrow \mathbb{R}^{+} \cup \{0\} $, an $n$-length block code with $k$-RS $(P_{S|X_{\cM}}, f, \varphi)$ will be required to satisfy the expected fidelity criterion ($d$, $\Delta$), i.e., 
\begin{align}
\begin{split}
 \label{eq:fidelity-criterion}
 \mathbb{E} \bigg [ d \bigg ( X_{\cM}^{n},  \varphi \big ( f (S^{n}, X_{S}^{n} ) \big ) \bigg ) \bigg ]  
 \triangleq  \mathbb{E} \bigg[ \dfrac{1}{n} \sum_{t=1}^{n} d \bigg ( X_{\cM t} , \Big ( \varphi \big ( f ( S^{n}, X_{S}^{n}) \big ) \Big )_{t}  \bigg ) \bigg ] \leq \Delta .
 \end{split}
\end{align}

We shall consider also the case where the decoder is informed of the sequence of sampled sets $S^{n}$. Denoting such an {\it informed decoder} by $\varphi_{S}$, the expected fidelity criterion \eqref{eq:fidelity-criterion} will use the augmented $\varphi_{S} \big( S^{n}, f(S^{n}, X_{S}^{n}) \big)$ instead of $\varphi \big( f(S^{n}, X_{S}^{n} ) \big).$ The earlier decoder (that is not informed) will be termed an {\it uninformed decoder}.

\begin{definition} \label{d:RDF}
 
A number $ R \geq 0$ is an achievable $k$-sample coding rate at average distortion level $\Delta$ if for every $\epsilon > 0$ and sufficiently large $n$, there exist $n$-length block codes with $k$-RS of rate less than $R + \epsilon$ and satisfying the expected fidelity criterion $(d, \Delta + \ep)$; and $(R, \Delta)$ will be termed an achievable $k$-sample rate distortion pair. The infimum of such achievable rates is denoted by $R^{I}( \Delta)$ for an informed decoder, and by $R^{U}( \Delta)$ for an uninformed decoder. We shall refer to $R^{I}(\Delta)$ as well as $R^{U}(\Delta)$ as the {\it sampling rate distortion function} (SRDf), suppressing the dependence on $k$. \end{definition}

\noindent {\it Remarks}: (i) Clearly, $R^{I}(\Delta) \leq R^{U}(\Delta),$ and both are nonincreasing in $k$. \\
(ii) For a DMMS $\{ X_{\cM t} \}_{t=1}^{\infty}$, the requirement \eqref{eq:k_MRS_def} on the sampler renders $\left \{ (X_{\cM t}, S_{t}) \right \}_{t=1}^{\infty}$ and thereby also $\left \{ (X_{S_{t}}, S_{t}) \right \}_{t=1}^{\infty}$ to be memoryless sequences. 



\section{Results}\label{s:Results}

Single-letter characterizations of the SRDfs in this paper involve, as an ingredient, a characterization of $R^{I}(\Delta)$ with $S_{t} = A, \ t= 1, \ldots , n$, where $A \subseteq {\cM}$ is a {\it fixed} set with $\left\vert{A}\right\vert = k$.  Denote the corresponding $R^{I} (\Delta)$ by $R_{A} ( \Delta)$ (with an abuse of notation). The {\it fixed-set} SRDf $R_{A}(\Delta),$ in effect, is the (standard) rate distortion function for the DMMS $\{ X_{A t } \}_{t=1}^{\infty}$ using a modified distortion measure $d_{A} : {\cal X}_{A} \times {\cal Y}_{\cM} \rightarrow \mathbbm{R}^{+} \cup \{ 0 \} $ defined by 
\begin{align} 
\label{eq:modified_distortion_A}
 d_{A}(x_{A}, y_{\cM}) = \mathbbm{E} [ d(X_{\cM}, y_{\cM}) | X_{A} = x_{A} ].
\end{align}

\begin{proposition} \label{th:k-FS-RD}
For a DMMS $\{ X_{\cM t} \}_{t=1}^{ \infty}$, the fixed-set SRDf for $A \subseteq {\cM}$ is 
\begin{align}
\label{eq:k-FS-RD}
R_{A} (\Delta) = 
\displaystyle \min_{X_{A^{c}} \MC X_{A} \MC Y_{\cM} \atop
 \mathbb{E} [d (X_{\cM}, Y_{\cM})] \leq {\Delta}}  I \big( X_{A} \wedge  Y_{\cM} \big) 
\end{align}
for $\Delta_{ \min,A} \leq \Delta \leq \Delta_{\max}$, and equals 0 for $\Delta \geq \Delta_{\max}$, where
\begin{align}
\begin{split}
\label{eq:Delta-max}
{\Delta}_{ \min, A} =  \mathbbm{E} \left [ \min_{y_{\cM} \in {\cal Y}_{\cM}} d_{A}(X_{A},y_{\cM}) \right ] , \ \ \ \ {\Delta}_{\rm max} = \min_{y_{\cM} \in {\cal Y}_{\cM}}  \mathbb{E} \big[ d (X_{\cM}, y_{\cM}) \big] =  \min_{y_{\cM} \in {\cal Y}_{\cM}} \mathbbm{E} [d_{A}(X_{A}, y_{\cM}) ].
\end{split}
\end{align}
\end{proposition}

\noeqref{eq:pe_distortion}

\begin{corollary} \label{cor:discrete-prob-error}
With ${\cal X}_{\cM} = {\cal Y}_{\cM}$, for the probability of error distortion measure
\begin{align}
\begin{split}
 \label{eq:pe_distortion}
 d(x_{\cM}, y_{\cM} ) = \mathbbm{1} ( x_{\cM} \neq y_{\cM} ) = 1 - \prod_{i=1}^{m}  \mathbbm{1}( x_{i} = y_{i} ), \ \ \ \ \  x_{\cM},  y_{\cM} \in {\cal X}_{\cM} 
\end{split}
 \end{align}
\noindent the SRDf is
\begin{align}
\label{eq:r_d_pe}
R_{A} (\Delta) = \begin{cases}
\displaystyle \min I \big( X_{A} \wedge Y_{A} \big), & \mbox{}{\Delta}_{ \min} \leq {\Delta} \leq {\Delta}_{\rm max} \\
0, & \mbox{}{\Delta} \geq {\Delta}_{\rm max},
\end{cases}
\end{align}
where the minimum in \eqref{eq:r_d_pe} is subject to 
\begin{equation}
 \label{eq:distortion_simplifed_discrete}
 \mathbbm{E}[ \alpha(X_{A}) \mathbbm{1}(X_{A} \neq Y_{A}) ] \leq \Delta - (1 - \mathbbm{E}[ \alpha(X_{A}) ] )
\end{equation}
with 
\begin{equation} \label{eq:alpha-proposition-fixed-set}
 \alpha(x_{A}) = \displaystyle \max_{ {\tilde x}_{A^{c}}  \in {\cal X}_{A^{c}} } P_{X_{A^{c}} | X_{A} } ( {\tilde x}_{A^{c}} |x_{A} )
\end{equation}
and 
\begin{align} \label{eq:Delta-min_pe_criterion}
\Delta_{ \min} = 1 -\mathbbm{E}[ \alpha(X_{A}) ] , \ \ \ \ \Delta_{ {\rm max}} = 1 - \max_{x_{\cM} \in {\cal X}_{\cM}}P_{X_{\cM}}(x_{\cM}).
\end{align}
\end{corollary}
\noindent {\it Remarks}: (i) The minimum in \eqref{eq:k-FS-RD} exists by virtue of the continuity in $P_{X_{\cM} Y_{\cM} }$ of $I(X_{A} \wedge Y_{\cM} ) $ over the compact set $\{ P_{X_{\cM} Y_{\cM}}:X_{A^{c}} \MC X_{A} \MC Y_{\cM}, \  \mathbbm{E}[ d(X_{\cM},Y_{\cM})] \leq \Delta   \}$. \\
(ii) The corollary relies on showing that the minimum in \eqref{eq:k-FS-RD} is attained now by a pmf $P_{X_{\cM} Y_{\cM}}$ under a longer Markov chain 
\begin{align}
X_{A^{c}} \MC X_{A} \MC Y_{A} \MC Y_{A^{c}} \label{eq:four_way_markov} .
\end{align}
Interestingly, the achievability proof entails in a first step a mapping of $x_{A}^{n}$ in ${\cal X}_{A}^{n} $ into its codeword $y_{A}^{n}$, from which in a second step a reconstruction $y_{A^{c}}^{n}$ of $x_{A^{c}}^{n}$ is obtained as a maximum a posteriori (MAP) estimate.

\vspace*{0.2cm}

The $k$-IRS affords a more capable mechanism than the fixed-set sampler of Proposition \ref{th:k-FS-RD}, with the sampling sets possibly varying in time. Surprisingly, the SRDf for a $k$-IRS, displayed as $R_{i}(\Delta)$, remains the same regardless of whether or not the decoder is provided information regarding the sequence of sampled sets. 

\begin{theorem}\label{th:k-IRS-informed-decoder}
For a $k$-IRS, the SRDf is
\begin{align}
\label{eq:k-IRS-informed-PMF}
R_{i}^{I} (\Delta) = R_{i}^{U}(\Delta) = R_{i}(\Delta) = \displaystyle \min  I \big( X_{S} \wedge Y_{\cM} | S \big)
\end{align}
for $\Delta_{\min} \leq \Delta \leq \Delta_{\max}$, where the minimum is with respect to $P_{X_{\cM} S Y_{\cM}} = P_{X_{\cM}} P_{S} P_{Y_{\cM}|S X_{S}}$ and $ \ \mathbbm{E} [d (X_{\cM}, Y_{\cM})] \leq {\Delta}$, with
\begin{align} \label{eq:Delta-min-IRS}
{\Delta}_{\min} =  \displaystyle \min_{ A \in {\cal A}_{k} }  \mathbbm{E} \left [ \min_{y_{\cM} \in {\cal Y}_{\cM}} d_{A}(X_{A}, y_{\cM}) \right ]
\end{align}
and $\Delta_{\max}$ as in \eqref{eq:Delta-max}.

\end{theorem}

A convenient equivalent expression for $R_{i}(\Delta)$ in \eqref{eq:k-IRS-informed-PMF} is given by
\begin{proposition} \label{prop:k-IRS-equiv}
For a $k$-IRS,
\begin{align} 
\label{eq:k-IRS-equiv}
 R_{i}(\Delta) = \min \sum \limits_{A \in \cAk} P_{S}(A) R_{A}(\Delta_{A}), \ \ \Delta_{\min} \leq \Delta \leq \Delta_{\max},
\end{align}
where the minimum is with respect to 
\begin{align}
 \label{eq:k-IRS-equiv-constraint-set}
 P_{S}, \ \Big \{ \Delta_{A} \geq \Delta_{\min, A}, \ A \in \cAk : \ \sum \limits_{A \in \cAk} P_{S}(A) \Delta_{A} \leq \Delta \Big \}.
\end{align}
\end{proposition}

\noindent {\it Proof.} For every $\Delta_{\min} \leq \Delta \leq  \Delta_{\max} $, in \eqref{eq:k-IRS-informed-PMF},
\begin{align}
\underset{ P_{X_{\cM}} P_{S} P_{Y_{\cM} |S X_{S}} \atop \mathbbm{E}[d(X_{\cM},Y_{\cM})] \leq \Delta } {\min} I(X_{S} \wedge Y_{\cM} | S ) & = \underset{ P_{X_{\cM}} P_{S} P_{Y_{\cM} |S X_{S}} \atop \mathbbm{E}[d(X_{\cM},Y_{\cM})] \leq \Delta } {\min} \sum_{A \in {\cal A}_{k}} P_{S}(A) I(X_{A} \wedge Y_{\cM}|S = A)  \\
   & = \underset{ P_{S}, \atop  \Delta_{A}:   \sum \limits_{A \in {\cal A}_{k}} P_{S} (A) \Delta_{A} \leq \Delta  } {\min}  \  \sum_{A \in {\cal A}_{k} } P_{S}(A) \underset{ P_{Y_{\cM} | S=A, X_{A}}   \atop \mathbbm{E}[d (X_{\cM},Y_{\cM}) | S = A] = \Delta_{A} } {\min} I(X_{A} \wedge Y_{\cM} | S = A) \label{eq:th1-ach-eq2} \\
    & = \underset{ P_{S}, \atop \Delta_{A}:    \sum \limits_{A \in {\cal A}_{k}} P_{S} (A) \Delta_{A} \leq \Delta } {\min} \ \sum_{A \in {\cal A}_{k} } P_{S}(A) R_{A}(\Delta_{A}),
\end{align}
where $P_{Y_{\cM} | S=A, X_{A}} $ is used to denote $P_{Y_{\cM} | S, X_{S}}( \cdot | A, \cdot) $ for compactness. The validity of \eqref{eq:th1-ach-eq2} follows by the introduction of the $\Delta_{A}$s and observing that the order of the minimization does not alter the value of the minimum. The last step obtains upon noting that the value of the inner minimum in  \eqref{eq:th1-ach-eq2} is the same upon replacing the equality in $\mathbbm{E}[d(X_{\cM},Y_{\cM}) | S = A ] = \Delta_{A}$  with ``$\leq$''. \qed

\vspace*{0.2cm}

\noindent {\it Remark}:
By Proposition \ref{prop:k-IRS-equiv}, the SRDf for a $k$-IRS is the lower convex envelope of the set of SRDfs $\{R_{A}(\Delta), \ A \in \cAk \}$ and thus is convex in $\Delta \geq \Delta_{\min}$. Furthermore,
\begin{align}
 R_{i}(\Delta) \leq \underset{ A \in \cAk} \min R_{A}(\Delta).
\end{align}
Additionally, a $k$-IRS can outperform strictly the best fixed-set sampler. For instance, if there is no fixed-set SRDf for any $A \in \cAk$ that is uniformly best for all $\Delta$, then the previous inequality can be strict. This is illustrated by the following example.

\vspace*{0.2cm}

\begin{example}
 With $\cM = \{1,2\}, \ {\cal X}_{1} = {\cal X}_{2} = {\cal Y}_{2} = \{0,1\}, $ and ${\cal Y}_{1} = \{0,1,e \},$ let $X_{1}, \ X_{2}$ be i.i.d. Bernoulli($0.5$) rvs, and 
 \begin{align}
  d \big ( (x_{1},x_{2}), (y_{1},y_{2}) \big ) = d_{1} (x_{1}, y_{1}) + d_{2}( x_{2}, y_{2})
 \end{align}
with
\begin{align}
 d_{1} (x_{1}, y_{1}) = \begin{cases}
                         0, \ \  & \text{if } x_{1} = y_{1} = 0; \ x_{1} = y_{1} = 1 \\
                         1, \ \  & \text{if } x_{1} = 0,  1, \ \ y_{1} = e \\
                         \infty, \ \ & \text{if } x_{1} = 0, \ y_{1} = 1; \ x_{1} = 1, \ y_{1} = 0,
                        \end{cases}
\end{align}
\vspace*{-0.3cm}
\begin{align}
 d_{2} (x_{2},y_{2}) = \mathbbm{1}(x_{2} \neq y_{2}).
\end{align}
For $k=1$,
\begin{align}
 R_{ \{1\} }(\Delta) & = 1.5 - \Delta, \ \ \ \ \ \ \ \  0.5 \leq \Delta \leq 1.5, \\
 R_{ \{2\} }(\Delta) & = 1 - h(\Delta-1), \ 1 \leq \Delta \leq 1.5
\end{align}
whereas
\begin{align}
 R_{i}(\Delta) = \begin{cases}
                         1.5515 - 1.103 \Delta,    & 0.5 \leq \Delta \leq 1.318, \\
                         1 - h(\Delta - 1),    &  1.318 \leq \Delta \leq 1.5,
                 \end{cases}
\end{align}
where $h( \cdot )$ is the binary entropy function.
Clearly, $R_{i}(\Delta)$ is strictly smaller than $\min \left \{R_{ \{1\}}(\Delta), R_{ \{2\} }(\Delta) \right \}$ for $ 0.5 <\Delta < 1.318;$ see Fig. \ref{fig:IRS_Fixed_set}. Note that while the distortion measure $d$ in Definition \ref{d:encoder} is taken to be finite-valued, the event $\left \{ d(X_{1}, Y_{1} )  = \infty  \right \}$ above is accommodated by assigning (optimally) zero probability to it.

\setlength{\unitlength}{0.8cm}
\begin{figure}[h]
\begin{center}
\includegraphics[
  width=12.5cm,height=8cm]{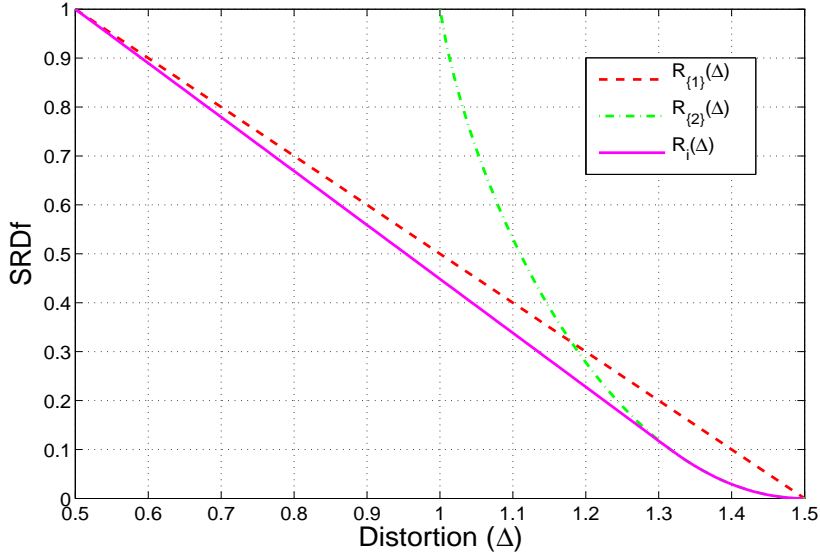}  
\end{center}
  \caption{SRDfs for $k$-IRS vs. fixed-set sampler}
  \label{fig:IRS_Fixed_set}
\end{figure}

\qed

\end{example}


A $k$-MRS is more powerful than a  $k$-IRS in that sampling with the former at each time instant can depend on the current DMMS realization. The SRDf for a $k$-MRS can improve with an informed decoder unlike for a $k$-IRS.

\begin{theorem} \label{th:k-MRS-informed-decoder}
For a $k$-MRS with informed decoder, the SRDf is 
\begin{align}
\label{eq:k-MRS-RD-informed}
\begin{split}
R_{m}^{I} (\Delta) = \min I \big( X_{S} \wedge Y_{\cM} | S, U \big) 
\end{split}
\end{align}
for $\Delta_{\min} \leq \Delta \leq \Delta_{\max}$, where the minimum is with respect to $ P_{U X_{\cM} S Y_{\cM}}  = P_{U} P_{X_{\cM}}  P_{S|X_{\cM} U} P_{Y_{\cM}|S X_{S} U}$ and $\mathbb{E} [d (X_{\cM}, Y_{\cM})] \leq {\Delta}$, with 
\begin{align}
 \label{eq:Delta-min-MRS}
&{\Delta}_{ \min} = \min_{P_{S|X_{\cM}} }  \mathbbm{E} \left [  \min_{P_{Y_{\cM}|S X_{S}}} \mathbbm{E}[ d(X_{\cM},Y_{\cM}) |S,X_{S}  ]   \right ] , 
\end{align} 
\begin{align} \label{eq:Delta-max-MRS}
{\Delta}_{ \max} = \min_{P_{S|X_{\cM}} }  \mathbbm{E} \left  [  \min_{ y_{\cM}} \ \mathbb{E}[ d(X_{\cM},y_{\cM}) |S  ]    \right ], \ \ \ 
\end{align}
and $U$ being a ${\cal U}$-valued rv with $|{\cal U}| \leq 3$.

\end{theorem}

\noindent {\it Remark}: Analogously as in Proposition \ref{prop:k-IRS-equiv}, the SRDf $R_{m}^{I}(\Delta)$ can be expressed as 
\begin{align}
 R_{m}^{I} (\Delta) = \underset{P_{U}, \ \Delta_{u}: \atop \sum \limits_{u} P_{U}(u) \Delta_{u} \leq \Delta } \min \sum \limits_{u} P_{U}(u) \underset{P_{S|X_{\cM}, U = u}, P_{Y_{\cM}|S X_{S}, U = u} \atop \mathbbm{E}[d(X_{\cM}, Y_{\cM}) | U = u] = \Delta_{u} } \min I(X_{S} \wedge Y_{\cM} | S, U = u) \label{eq:k-MRS_equiv_form}
\end{align}
and thereby equals a lower convex envelope of functions of $\Delta$.

\vspace*{0.1cm}

The optimal sampler that  attains the SRDf in Theorem \ref{th:k-MRS-informed-decoder} has a simple structure. It is easy to see that each of $\Delta_{\min}$ and $\Delta_{\max}$ in \eqref{eq:Delta-min-MRS} and \eqref{eq:Delta-max-MRS}, respectively, is attained by a sampler for which $P_{S|X_{\cM}}$ takes the form of a conditional point-mass. Such samplers, in fact, are optimal for every distortion level $ \Delta_{\min} \leq \Delta \leq \Delta_{\max}$ and will depend on $\Delta$, in general.

\begin{definition} \label{d:point_mass_sampler}
 Given a mapping $h : {\cal X}_{\cM} \times {\cal U} \rightarrow {\cal A}_{k}$,  the (conditional point-mass) mapping $\delta_{  h( \cdot ) }: {\cal A}_{k} \rightarrow \{0,1\} $ is defined by
 \begin{equation}
 \label{eq:point_mass_sampler}
 \delta_{  h(x_{\cM}, u) }(s)   \triangleq \begin{cases}
                                1, \  & s = h(x_{\cM}, u)  \\
                                0, \  & \text{otherwise}, \hspace*{0.5cm} \  (x_{\cM},u) \in {\cal X}_{\cM} \times {\cal U}, \ s \in \cAk. 
                               \end{cases}
 \end{equation}
\end{definition}

\noindent The following reduction of Theorem \ref{th:k-MRS-informed-decoder} shows the optimality of conditional point-mass samplers for a $k$-MRS which will be seen to play a material role in the achievability proof of Theorem \ref{th:k-MRS-informed-decoder}.

\begin{theorem} \label{th:k-MRS-informed-alternative}
 For a $k$-MRS with informed decoder, the SRDf equals
 \begin{align}
 \label{eq:k-MRS-informed-alternative}
R_{m}^{I}(\Delta) =  \min I \big( X_{S} \wedge Y_{\cM} | S, U \big) 
 \end{align}
for $ \Delta_{\min} \leq \Delta \leq \Delta_{\max},$ with $\Delta_{\min}$ and $\Delta_{\max}$ as in \eqref{eq:Delta-min-MRS} and \eqref{eq:Delta-max-MRS}, respectively, where the minimum is with respect to  $ P_{U X_{\cM} S  Y_{\cM}}$ of the form  $P_{U}  P_{X_{\cM}} \delta_{  h( \cdot) } P_{Y_{\cM}|S X_{S} U}$ with $\mathbb{E} [d (X_{\cM}, Y_{\cM})]  \leq {\Delta},$ where the (time-sharing) rv $U$ takes values in ${\cal U}$ with $|{\cal U}| \leq 3 $. 
\end{theorem}

The structure of the optimal sampler in Theorem \ref{th:k-MRS-informed-alternative} implies that the search space for minimization now can be reduced to the corner points of the simplexes of the conditional pmfs $P_{S| X_{\cM} U}( \cdot | x_{\cM}, u ), \ (x_{\cM},u) \in {\cal X}_{\cM} \times \cU $. The SRDf in \eqref{eq:k-MRS-informed-alternative} is thus the lower  convex envelope of the SRDfs for conditional point-mass samplers. In general, time-sharing between such samplers will be seen to achieve the best compression rate for a given distortion level.

\vspace*{0.2cm}


Finally, for a $k$-MRS with uninformed decoder, we provide an upper bound for the SRDf.

\begin{theorem} \label{th:k-MRS-uninformed-decoder}
For a  $k$-MRS  with uninformed decoder,  
\begin{align} \label{eq:k-MRS-RD-uninformed}
 R_{m}^{U} (\Delta) 
\displaystyle \leq \min I \big( S, X_{S} \wedge Y_{\cM} \big)
\end{align}
for $\Delta_{\min} \leq \Delta \leq \Delta_{\max}$, where the minimum is with respect to $P_{X_{\cM}SY_{\cM}} = P_{X_{\cM}} P_{S|X_{\cM}} P_{Y_{\cM}|S X_{S}} $ and \\ $\mathbbm{E} [ d(X_{\cM}, Y_{\cM}) ] \leq \Delta,$ with $\Delta_{\min} \ { and} \ \Delta_{\max}$ being as in (\ref{eq:Delta-min-MRS}) and \eqref{eq:Delta-max}. 
\end{theorem}

The (achievability) proof of Theorem \ref{th:k-MRS-uninformed-decoder} is along the lines of Proposition \ref{th:k-FS-RD}. The lack of a  converse is due to the inability to prove or disprove the convexity of the right-side of \eqref{eq:k-MRS-RD-uninformed} in $\Delta$. (Convexity would imply equality in \eqref{eq:k-MRS-RD-uninformed}.)
The optimal sampler can, however, be shown to be a conditional point-mass sampler \eqref{eq:point_mass_sampler} along the lines of Theorem \ref{th:k-MRS-informed-alternative}.
Note that the same conditional point-mass sampler need not be the best in \eqref{eq:k-MRS-RD-informed} and \eqref{eq:k-MRS-RD-uninformed}.

\vspace*{0.2cm}

Strong forms of the $k$-MRS and $k$-IRS are obtained by allowing time-dependence in sampling. Specifically, \eqref{eq:k_MRS_def} and \eqref{eq:k_IRS_def} can be strengthened, respectively, to 
\begin{align}
 \label{eq:k_SMRS_def}
P_{S_{t}|X_{\cM}^{t} S^{t-1}} = P_{S_{t}|X_{\cM t} S^{t-1}}
\end{align}
and 
\begin{align}
 \label{eq:k_SIRS_def}
P_{S_{t}|X_{\cM}^{t} S^{t-1}} = P_{S_{t}|S^{t-1}}.
\end{align}
Surprisingly, this does not improve SRDf for the $k$-MRS (with decoder informed) or the $k$-IRS. 

\begin{proposition}
 \label{prop:SMRS_SIRS}
 For a strong $k$-MRS in \eqref{eq:k_SMRS_def} and a strong $k$-IRS in \eqref{eq:k_SIRS_def}, the corresponding SRDfs $R_{ms}^{I}(\Delta)$  and $R_{is}(\Delta)$ equal the right-sides of \eqref{eq:k-MRS-RD-informed}  and \eqref{eq:k-IRS-informed-PMF}, respectively.
\end{proposition}

Finally, standard properties of the SRDf for the fixed-set sampler, $k$-IRS and $k$-MRS with informed decoder are summarized in the following
\begin{lemma}
\label{l:SRDf_convexity}
For a fixed $P_{X_{\cM}}$, the right-sides of \eqref{eq:k-FS-RD}, \eqref{eq:k-IRS-informed-PMF} and \eqref{eq:k-MRS-RD-informed}  are finite-valued, nonincreasing, convex, continuous functions of $\Delta$.
\end{lemma}

We close this section with an example showing that (i) the SRDf for a $k$-MRS with informed decoder can be strictly smaller than that of a $k$-IRS; and (ii) furthermore, unlike for a $k$-IRS,  a $k$-MRS with informed decoder can outperform strictly that with an uninformed decoder, uniformly for all feasible distortion values.
\begin{example}
 With $\cM = \{1,2\}$ and ${\cal X}_{1} = {\cal X}_{2} = \{ 0,1 \},$ consider a DMMS with $P_{X_{1} X_{2}}$ represented by a virtual binary symmetric channel (BSC) shown in Figure \ref{fig:Virtual_BSC}.
  \setlength{\unitlength}{0.8cm}
\begin{figure}[h]
\begin{center}
\includegraphics[
  width=5.5cm,height=3cm]{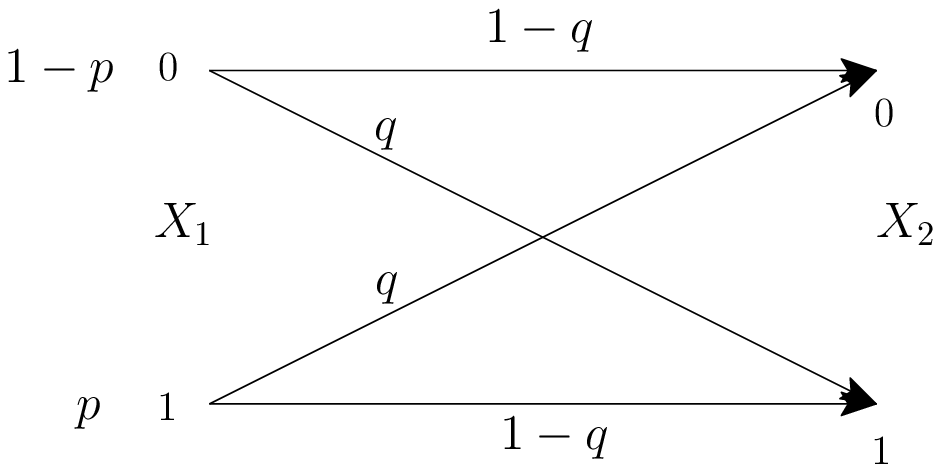}  
\end{center}
  \caption{ BSC ($q$)}
  \label{fig:Virtual_BSC}
\end{figure}
Fix $p \leq 0.5$ and $q = 0.5,$ i.e., $X_{1}$ and $X_{2}$ are independent. Let $d$ correspond to the probability of error criterion, i.e., $d(x_{\cM}, y_{\cM}) = \mathbbm{1} (x_{\cM} \neq y_{\cM}).$

\noindent (i) Considering a $k$-MRS, $k = 1$, with informed decoder, we obtain by Theorem \ref{th:k-MRS-informed-alternative} that $\Delta_{\min} = 0, \ \Delta_{\max} = p,$ and the (conditional point-mass) sampler  

\begin{align}
\label{eq:Optimal_sampler}
 P_{S|X_{\cM}}(s|x_{\cM}) = \begin{cases}
                             1, \ \ & s = 1, \ x_{\cM} = 00  \ \text{or} \ 11 \\
                             1, \ \ & s = 2, \ x_{\cM} = 01  \ \text{or} \ 10 \\
                             0, \ \ & \text{otherwise}
                            \end{cases}
\end{align}
is uniformly optimal for all $0 \leq \Delta \leq p,$ and 
\begin{align}
 R_{m}^{I}(\Delta) = h(p) - h(\Delta), \ \ \ 0 \leq \Delta \leq p.
\end{align}
To obtain $R_{i}(\Delta),$   the SRDfs for fixed-set samplers \eqref{eq:k-FS-RD} are 
 \begin{align}
  R_{ \{1 \} }( \Delta) = h(p) - h \left (   2\Delta - 1    \right ), \ \  \tfrac{1}{2} \leq \Delta \leq \frac{ 1 + p}{2},
 \end{align}
 and 
\begin{align}
  R_{ \{2 \} }( \Delta) = h \left ( \frac{1}{2} \right ) - h \left ( \frac{ \Delta - p }{ 1 - p}  \right ), \ \  p \leq \Delta \leq \frac{ 1 + p}{2}.
 \end{align}
Since $R_{ \{2\} } (\Delta) \leq R_{ \{1\} } (\Delta)$ uniformly in $\Delta,$ it is a simple exercise to show that 
\begin{align}
 R_{i}(\Delta) = R_{ \{2\} }(\Delta) .
\end{align}
Clearly, $R_{m}^{I}(\Delta) \leq R_{i}(\Delta)$, with $\Delta_{\max}$ for the former being $\Delta_{\min}$ for the latter, as shown in Figure \ref{fig:MRS_IRS}.

\begin{figure}[h]
\centering
\includegraphics[width=7.5cm,height=6cm]{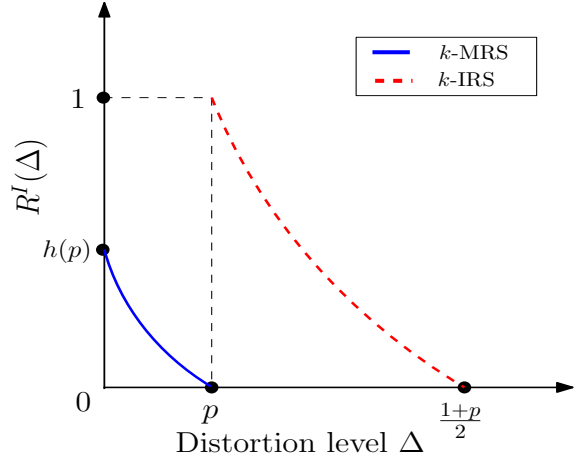}
  \caption{SRDf for $k$-MRS vs. $k$-IRS}
  \label{fig:MRS_IRS}    
  \vspace*{-0.15 cm}
\end{figure}

\noindent (ii) The conditional pmf $P_{S|X_{\cM}}$ in \eqref{eq:Optimal_sampler} represents a $1$-$1$ map between the values of $X_{\cM}$ and $(S,X_{S})$, and can be seen also to be the optimal choice in the right-side of \eqref{eq:k-MRS-RD-uninformed} for all $0 \leq \Delta \leq \frac{1+p}{2}.$ The remaining minimization in \eqref{eq:k-MRS-RD-uninformed}, with respect to $P_{Y_{\cM}|S X_{S}}$, renders the right-side to be convex in $\Delta.$ Consequently, as observed in the passage following Theorem \ref{th:k-MRS-uninformed-decoder}, the bound in \eqref{eq:k-MRS-RD-uninformed} is tight. The resulting values of $R_{m}^{I}(\Delta)$ and $R_{m}^{U}(\Delta)$ are plotted for $p=0.1$ in Figure \ref{fig:MRS_Inf_vs_Uninf}. \\

\begin{figure}[h]
\centering
\includegraphics[width=10.3cm,height=7.0cm]{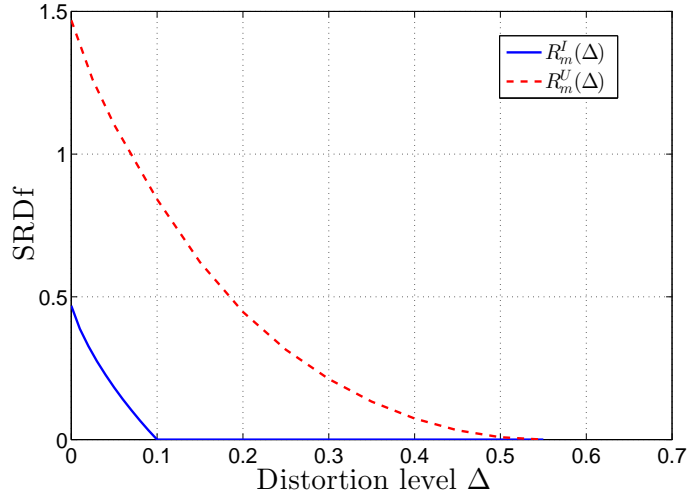}
  \caption{SRDf for $k$-MRS}
  \label{fig:MRS_Inf_vs_Uninf}
\end{figure}
\end{example}






\section{Proofs} \label{s:Proofs}

\subsection{Achievability proofs}
Our achievability proofs successively build upon each other in the order: fixed-set sampler, $k$-IRS and $k$-MRS. The achievability proof of Proposition \ref{th:k-FS-RD} for a fixed-set sampler forms a basic building block for subsequent application. Relying on this, the SRDf for a $k$-IRS is shown to be achieved in Theorem \ref{th:k-IRS-informed-decoder} without the decoder being informed of the sequence of sampled sets. Next, for a $k$-MRS with informed decoder, we prove first Theorem \ref{th:k-MRS-informed-alternative} which shows that the optimal sampler is deterministic in that the corresponding $P_{S |X_{\cM }}$ is a point-mass. This structure enables an achievability proof of Theorem \ref{th:k-MRS-informed-decoder} which builds on that of Proposition \ref{th:k-FS-RD}. Lastly, for a $k$-MRS with uninformed decoder, the achievability proof of Theorem \ref{th:k-MRS-uninformed-decoder} rests on the preceding proofs.

\vspace*{0.3cm}

\noindent {\bf  Proposition \ref{th:k-FS-RD}}: Observe first that
\begin{align}
 \Delta_{\min, A} & = \underset{ X_{A^{c}} \MC X_{A} \MC Y_{\cM} } \min \mathbbm{E}[d(X_{\cM}, Y_{\cM})] \\
 & = \underset{ X_{A^{c}} \MC X_{A} \MC Y_{\cM} } \min \mathbbm{E} \big [ \mathbbm{E}[d(X_{\cM}, Y_{\cM})|X_{A}] \big ] \\
 & = \underset{P_{ X_{A} Y_{\cM}} } \min \mathbbm{E}[d_{A}(X_{A}, Y_{\cM})]  \ \ \ \text{by } \eqref{eq:modified_distortion_A} \text{ and since } X_{A^{c}} \MC X_{A} \MC Y_{\cM} \\
 & =   \mathbbm{E} \Big [ \  \underset{y_{\cM} \in \cYm } \min d_{A}(X_{A}, y_{\cM}) \Big ]
\end{align}
and
\begin{align}
 \Delta_{\max} & = \underset{X_{A^{c}} \MC X_{A} \MC Y_{\cM} \atop P_{X_{A} Y_{\cM}} = P_{X_{A}} P_{Y_{\cM}}  } \min \mathbbm{E}[d(X_{\cM}, Y_{\cM})] \\
 & = \underset{ P_{X_{\cM}} P_{Y_{\cM}}  } \min \mathbbm{E}[d(X_{\cM}, Y_{\cM})] \\
 & = \underset{ y_{\cM} \in \cYm  } \min \mathbbm{E}[d(X_{\cM}, y_{\cM})].
\end{align}

Next, note that for every $\Delta_{\min, A} \leq \Delta \leq {\Delta}_{\max}$,
\begin{align} \label{eq:prop1-equiv-form}
\underset{X_{A^{c}} \MC X_{A} \MC Y_{\cM} \atop \mathbbm{E}[ d (X_{\cM},Y_{\cM}) ] \leq \Delta} {\min} I(X_{A} \wedge Y_{\cM}) = \underset{ \mathbbm{E}[ {d}_{A}( X_{A},Y_{\cM} ) ] \leq \Delta} {\min} I(X_{A} \wedge Y_{\cM} ) .
\end{align}

\vspace*{0.1cm}

Clearly every feasible $P_{X_{\cM} Y_{\cM}} = P_{X_{A^{c}} X_{A} Y_{\cM}}$ on the left-side above gives a feasible $P_{X_{A} Y_{\cM}}$ on the right-side. Similarly every feasible $P_{X_{A} Y_{\cM}}$ on the right-side leads to a feasible $P_{X_{\cM} Y_{\cM}}$ on the left-side of the form $P_{X_{\cM} Y_{\cM}} = P_{X_{A^{c}}| X_{A}} P_{X_{A} Y_{\cM}}$.
\vspace*{0.1cm}

Given $\ep > 0$, consider a (standard) rate distortion code $(f, \varphi)$ for the DMMS $\{ X_{A t} \}_{t=1}^{\infty}$ with distortion measure ${d}_{A}$, of rate $ \frac{1}{n} \log || f||  \leq R_{A}(\Delta) + \epsilon$ and with expected distortion $\mathbbm{E} \left [ {d}_{A} \big (X_{A}^{n}, \varphi(f(X_{A}^{n})) \big ) \right ] \leq \Delta + \ep $ for all $n \geq N_{A}(\ep),$ say.

\vspace*{0.1cm}

The code $(f, \varphi)$ also satisfies 
\begin{align}
 \mathbbm{E}[d (X_{\cM}^{n}, Y_{\cM}^{n}) ] &= \dfrac{1}{n} \mathbbm{E} \left [ \sum_{t=1}^{n} \mathbbm{E} \Big[ d \Big (X_{\cM t}, \big ( \varphi(f(X_{A}^{n})) \big )_{t} \Big)\Big | X_{A}^{n} \Big] \right]  \\ 
 & =  \dfrac{1}{n} \mathbbm{E} \left [ \sum_{t=1}^{n} \mathbbm{E} \Big[ d \Big (X_{\cM t}, \big ( \varphi(f(X_{A}^{n})) \big )_{t} \Big)\Big | X_{A t} \Big] \right]  \\
 & = \mathbbm{E} \left [ {d}_{A} \Big (X_{A}^{n}, \varphi \big (f(X_{A}^{n}) \big ) \Big ) \right ]  \\
 & \leq \Delta + \ep,
\end{align}
thereby yielding achievability in the proposition.


\vspace*{0.3cm}

Turning to the corollary, for every $P_{X_{\cM}Y_{\cM}}$ satisfying the constraints in \eqref{eq:k-FS-RD}, consider the pmf $Q_{X_{\cM}Y_{\cM} }$ defined by 
 \begin{align}
  Q_{X_{\cM}Y_{\cM}}(x_{\cM},y_{\cM}) \triangleq P_{X_{\cM} Y_{A}}(x_{\cM},y_{A}) \mathbbm{1} \big ( y_{A^{c}} = MAP \ (y_{A}) \big ) , \ \ x_{\cM}, y_{\cM} \in {\cal X}_{\cM}, \label{eq:cor1-eq1} 
 \end{align}
where 
\begin{align}
MAP \ (y_{A}) = \underset{ {\tilde y}_{A^{c}} \in {\cal Y}_{A^{c}} } { \arg \ \max} \  P_{X_{A^{c}}|X_{A}}({\tilde y}_{A^{c}}|y_{A}) \label{eq:MAP_def_cor} 
\end{align}
is the maximum a posteriori estimate of $y_{A^{c}}$ given   $y_{A}$ according to $P_{X_{A^{c}}|X_{A}}$. Observe that $Q_{X_{\cM} Y_{\cM}}$ satisfies
\begin{align}
\label{eq:four_way_markov_prf_cor} 
 Q_{X_{A^{c}}} \MC Q_{X_{A}} \MC Q_{Y_{A}} \MC Q_{Y_{A^{c}}}
\end{align}
and 
\begin{align}
 \mathbbm{E}_{P}[d(X_{\cM}, Y_{\cM})]  & = P(X_{\cM} \neq Y_{\cM}) \\
 & = P(X_{A} \neq Y_{A}) + P(X_{A} = Y_{A}) P(X_{A^{c}} \neq Y_{A^{c}}|X_{A} = Y_{A}) \\
 & = Q(X_{A} \neq Y_{A}) + Q(X_{A} = Y_{A}) P(X_{A^{c}} \neq Y_{A^{c}}|X_{A} = Y_{A}) \\
 & \geq Q(X_{A} \neq Y_{A}) + Q(X_{A} = Y_{A}) Q(X_{A^{c}} \neq Y_{A^{c}}|X_{A} = Y_{A}) \\
 & = Q(X_{\cM} \neq Y_{\cM}) = \mathbbm{E}_{Q}[d(X_{\cM}, Y_{\cM})], \label{eq:cor_prf_distortion}
\end{align}
where the inequality is by \eqref{eq:cor1-eq1}, \eqref{eq:MAP_def_cor} and the optimality of the MAP estimator. Also, it is readily checked that
\begin{align}
\label{eq:cor_distortion_alpha}
 \mathbbm{E}_{Q}[d(X_{\cM}, Y_{\cM})] = 1 - \mathbbm{E}[ \alpha(X_{A})] + \mathbbm{E}[ \alpha(X_{A}) \mathbbm{1}(X_{A} \neq Y_{A}) ].
\end{align}
Furthermore,
\begin{align}
\label{eq:cor_mutual_info}
 I_{Q}( X_{A} \wedge Y_{\cM} ) = I_{Q}( X_{A} \wedge Y_{A} ) = I_{P}( X_{A} \wedge Y_{A} ) \leq I_{P}( X_{A} \wedge Y_{\cM} ).
\end{align}
Putting together  \eqref{eq:cor1-eq1}  - \eqref{eq:cor_mutual_info} and comparing with \eqref{eq:k-FS-RD}  establishes the corollary.

\noeqref{eq:cor1-eq1} \noeqref{eq:MAP_def_cor}, \noeqref{eq:four_way_markov_prf_cor} \noeqref{eq:cor_prf_distortion} \noeqref{eq:cor_mutual_info}

\vspace*{0.2cm}

It is interesting to note that the form of \eqref{eq:k-FS-RD} 
\begin{align} 
 \min_{X_{A^{c}} \MC X_{A} \MC Y_{\cM} \atop P(X_{\cM} \neq Y_{\cM}) \leq \Delta }  I(X_{A} \wedge Y_{\cM}) 
  =  \min_{  \mathbbm{E}[ \alpha(X_{A}) \mathbbm{1}(X_{A} \neq Y_{A}) ] \leq \Delta - (1 - \mathbbm{E}[ \alpha(X_{A})] ) }I(X_{A} \wedge Y_{A} )
 \end{align}
leads to a simpler and direct proof of achievability of the corollary. Specifically, for a given $\Delta$, first $x_{A}^{n}$ is mapped into (only) its corresponding codeword $y_{A}^{n}$ but under a modified distortion measure ${\tilde d}(x_{A}, y_{A}) \triangleq \alpha(x_{A}) \mathbbm{1}(x_{A} \neq y_{A}) $ and a corresponding reduced threshold as indicated by \eqref{eq:distortion_simplifed_discrete}. Next, the codewords $y_{A}^{n}$ serve as sufficient statistics from which (the unsampled) $x_{A^{c}}^{n}$ is reconstructed as $y_{A^{c}}^{n} = MAP \ (y_{A}^{n})$ under $P_{X_{A^{c}}^{n}|X_{A}^{n}};$ the corresponding estimation error coincides with the reduction in the threshold.

\qed


\noindent {\bf  Theorem \ref{th:k-IRS-informed-decoder}}:  The equivalent expression for $R_{i}(\Delta)$ given by Proposition \ref{prop:k-IRS-equiv} suggests an achievability scheme using a concatenation of fixed-set sampling rate distortion codes from Proposition \ref{th:k-FS-RD}. Let $P_{S}$ and $\{ \Delta_{A}, \  A \in \cAk \}$ yield the minimum in Proposition \ref{prop:k-IRS-equiv}. A sequence of sampling sets $S^{n}$ are constructed {\it a priori} with $S_{t} = A$ repeatedly for approximately $nP_{S}(A)$ time instants, for each $A$ in $\cAk.$ Correspondingly, sampling rate distortion codes of blocklength $\cong n P_{S}(A)$ -- with distortion $\cong \Delta_{A}$ and of rate $ \cong R_{A}(\Delta_{A})$ -- are concatenated. This predetermined selection of sampling sets does not require the decoder to be additionally informed.

\vspace*{0.2cm}

For a fixed $\Delta_{\min} \leq \Delta \leq \Delta_{\max}$, let $P_{S}$ and $\{ \Delta_{A}, \ A \in {\cal A}_{k} \} $ attain the minimum in \eqref{eq:k-IRS-equiv}. Fix $ \ep >0 $ and  $ 0 < \ep' < \ep.$
Order (in any manner) the elements of ${\cal A}_{k}$ as $A_{i}, \ i \in {\cal M}_{k} \triangleq \{ 1 , \ldots, M_{k} \},$ with $ M_{k} = {m \choose k}.$ For $i \in {\cal M}_{k} $ and $n \geq 1$, define the ``time-sets'' $\nu_{A_{i}}$ as
\begin{align}
 \nu_{A_{i}} = \Bigg \{t : \lceil n \sum_{j=1}^{i-1} P_{S}(A_{j}) \rceil + 1 \leq t  \leq \lceil n \sum_{j=1}^{i} P_{S}(A_{j}) \rceil  \Bigg   \}, \ \ \ A_{i} \in \cAk.
\end{align}
The time-sets cover $\{ 1, \ldots, n \}$, i.e.,
 \begin{align}
  \underset{i \in  {\cal M}_{k} } { \bigcup } \nu_{A_{i}} = \{ 1, \ldots, n \}
 \end{align}
and satisfy
 \begin{align}
   \left |  \dfrac{|\nu_{A_{i}}|}{n} - P_{S}(A_{i}) \right | \leq \frac{1}{n}, \ \ \ i \in {\cal M}_{k}.
 \end{align}
 
Now, a $k$-IRS is chosen with a deterministic sampling sequence $S^{n} = s^{n}$ according to
\begin{align}
 S_{t} = s_{t} = A_{i},  \ \ t \in \nu_{A_{i}}, \ A_{i} \in \cAk. 
\end{align}
By Proposition \ref{th:k-FS-RD}, for each $  A_{i}$ in $ \cAk$, there exists a code $(f_{A_{i}}, \varphi_{A_{i}})$, $f_{A_{i}}: {\cal X}_{A_{i}}^{\nu_{A_{i}}} \rightarrow \{1, \ldots, J_{A_{i}} \}$ and $\varphi_{A_{i}}: \{1, \ldots, J_{A_{i}} \} \rightarrow {\cal Y}_{\cM}^{ \nu_{A_{i}} }$ of rate $ \dfrac{1}{|\nu_{A_{i}}|} \log J_{A_{i}} \leq R_{A_{i}} (\Delta_{A_{i}}) + \frac{\ep'}{2}  $ and with 
\begin{align}
\mathbbm{E} \left [ d \Big ( X_{\cal M}^{\nu_{A_{i}}}, \varphi_{A_{i}} \big ( f_{A_{i}}(X_{A_{i}}^{\nu_{A_{i}}}) \big ) \Big )  \right ] = \mathbbm{E} \left [ d_{A_{i}} \Big ( X_{A_{i}}^{\nu_{A_{i}}}, \varphi_{A_{i}} \big ( f_{A_{i}}(X_{A_{i}}^{\nu_{A_{i}}}) \big ) \Big )  \right ]   \leq \Delta_{A_{i}} + \frac{\ep'}{2} 
\end{align}
for all $|\nu_{A_{i}}| \geq N_{A_{i}} \left (\frac{\ep'}{2} \right)$ (cf. proof of Proposition \ref{th:k-FS-RD}).

\vspace*{0.2cm}

Consider a (composite) code $(f, \varphi)$  as follows. For the deterministic sampling scheme defined above, the encoder $f$ consists of a concatenation of encoders defined by
\begin{align}
f(S^{n}, x^{n} ) =  \left ( f_{A_{1}} \left (x_{A_{1}}^{\nu_{A_{1}}} \right ), \ldots, f_{A_{M_{k}}} \left(x_{A_{ M_{k} }}^{\nu_{A_{ M_{k} }}} \right ) \right), \ \ \ x^{n} \in  \mathop{\mbox{\large $\times$}} \limits_{i = 1}^{ M_{k}}   {\cal X}_{A_{i}}^{\nu_{A_{i}}}
\end{align}
which maps the output of the $k$-IRS into the set ${\cal J} \triangleq  \mathop{\mbox{\large $\times$}} \limits_{i = 1}^{ M_{k}}  \left \{1 , \ldots, J_{A_{i}} \right \} $. The decoder $\varphi$ is given by
\begin{align}
 \varphi \big ( j_{1}, \ldots, j_{M_{k}} \big ) \triangleq  \left (  \varphi_{A_{1}} \left( j_{1} \right), \ldots, \varphi_{A_{M_{k}}} \left( j_{M_{k}} \right) \right ), \ \ \ \ (j_{1}, \ldots, j_{M_{k}} ) \in {\cal J},
\end{align}
and is aware of the sampling sequence without being  informed additionally of it.

\vspace*{0.2cm}

The rate of the code is
\begin{align}
\dfrac{1}{n} \log || \cJ || & = \dfrac{1}{n} \sum \limits_{i = 1}^{M_{k}} \log J_{A_{i}} \\
& \leq  \dfrac{1}{n} \left  ( \sum \limits_{i = 1}^{M_{k}}  |\nu_{A_{i}}| \left (R_{A_{i}}(\Delta_{A_{i}})+ \frac{\ep'}{2} \right )  \right ) \\
& \leq   \sum \limits_{i = 1}^{M_{k}} \left ( \Big (P_{S}(A_{i}) + \frac{1}{n} \Big ) \left (R_{A_{i}}(\Delta_{A_{i}})+ \frac{\ep'}{2} \right )   \right )  \\
& \leq \sum \limits_{i = 1}^{M_{k}} P_{S}(A_{i}) R_{A_{i}}(\Delta_{A_{i}}) + \ep' < R_{i}(\Delta) + \ep \label{eq:Rate_eq1},
\end{align}
where the previous inequality holds for all $n$ large enough.
 Denoting the decoder output by $Y_{\cM}^{n} \triangleq \varphi \left ( f(S^{n}, X_{S}^{n} ) \right ) $, we have that
\begin{align}
 \mathbbm{E}[ d(X_{\cM}^{n},Y_{\cM}^{n})] & = \mathbbm{E} \left [  \dfrac{1}{n} \sum_{t=1}^{n} d(X_{\cM t},Y_{\cM t})  \right ] \\ 					   
					  & = \dfrac{1}{n}  \sum \limits_{ i  = 1}^{M_{k}} |\nu_{A_{i}}|  \mathbbm{E} \left [ d \left ( X_{\cal M}^{\nu_{A_{i}}}, \varphi_{A_{i}} \left ( f_{A_{i}}(X_{A_{i}}^{\nu_{A_{i}}}) \right ) \right )  \right ] \\ 
					  & \leq \sum \limits_{ i  = 1}^{M_{k}} \left (P_{S}(A_{i}) + \frac{1}{n} \right ) \left (\Delta_{A_{i}} + \frac{\ep'}{2} \right ) \\
					  & = \sum \limits_{i=1}^{M_{k}} P_{S}(A_{i}) \Delta_{A_{i}} + \frac{M_{k}}{n} \left ( \frac{\ep'}{2} + \Delta_{\max} \right ) + \frac{\ep'}{2} \\
					  & \leq \Delta + \ep \label{eq:Distortion_eq1}
\end{align}
by \eqref{eq:k-IRS-equiv-constraint-set} and for all $n$ large enough. The proof is completed by noting that \eqref{eq:Rate_eq1} and \eqref{eq:Distortion_eq1} hold simultaneously for all $n$ large enough.
\qed
 
 
 
\vspace*{0.5cm} 


Next, we establish Theorem \ref{th:k-MRS-informed-alternative}. The structure of the conditional point-mass sampler therein will be used next in the achievability proof of Theorem \ref{th:k-MRS-informed-decoder} to follow.

\vspace*{0.2cm}

\noindent {\bf Theorem \ref{th:k-MRS-informed-alternative}}: 
Denoting the minima in \eqref{eq:k-MRS-RD-informed} and \eqref{eq:k-MRS-informed-alternative} by $q(\Delta)$ and $r(\Delta)$, respectively, clearly
\begin{align}
 q(\Delta) \leq r(\Delta), \ \ \ \ \ \Delta_{\min} \leq \Delta \leq \Delta_{\max}.
 \label{eq:Alternative_ach_eq_compare}
\end{align}
In fact, equality will be shown to hold, thereby proving the theorem. First, since $q(\Delta)$ and $r(\Delta)$ are convex in $\Delta$ by Lemma \ref{l:SRDf_convexity}, by \cite[Lemma 8.1]{CsiKor11} they can be expressed in terms of their Lagrangians as 
\begin{align}
  q(\Delta) = \underset{\lambda \geq 0} \max \   G_{q}(\lambda) - \lambda \Delta   \ \ \ \text{and }  \ \ r(\Delta) = \underset{\lambda \geq 0} \max \  G_{r}(\lambda) - \lambda \Delta  ,
 \label{eq:Lagrangian_exp_SRDf}
\end{align}
where $G_{q}(\lambda)$ and $G_{r}(\lambda)$ are the respective minima of
\begin{align}
 I(X_{S} \wedge Y_{\cM} |S, U) + \lambda \mathbbm{E}\left[ d(X_{\cM}, Y_{\cM}) \right]
 \label{eq:Alternative_ach_eq1}
\end{align}
over $\left ( P_{U}, P_{S|X_{\cM} U }, P_{Y_{\cM}|S X_{S} U}  \right)$ and $\left (P_{U}, \delta_{ h(\cdot) }, P_{Y_{\cM}|S X_{S} U}  \right)$. By the conditional version of Tops\o e's identity \cite[Lemma 8.5]{CsiKor11}, the expression in \eqref{eq:Alternative_ach_eq1} equals
\begin{align}
\underset{ Q_{Y_{\cM}|SU} } \min D \left ( P_{Y_{\cM}|SX_{S} U} \big | \big | Q_{Y_{\cM}| S U } \big | P_{ S X_{S} U} \right ) + \lambda \mathbbm{E}\left[ d(X_{\cM}, Y_{\cM}) \right].
 \label{eq:Alternative_ach_eq2}
\end{align}

In $G_{q}(\lambda)$, the minimum of the expression in \eqref{eq:Alternative_ach_eq2} also over  $\left (P_{U}, P_{S|X_{\cM} U }, P_{Y_{\cM}|S X_{S} U}   \right)$ is not altered by changing the order of minimization with $P_{S|X_{\cM} U}$ being the innermost. Using this fact, it is shown in  Appendix that the minimizing $P_{S|X_{\cM} U}$ is of the form $\delta_{h(\cdot)}$, whereby 
\begin{align}
G_{q}(\lambda) = G_{r}(\lambda).  \label{eq:Alternative_ach_eq4}
\end{align}
Hence, equality holds in \eqref{eq:Alternative_ach_eq_compare}.

\qed

\vspace*{0.2cm}


\noindent {\bf  Theorem \ref{th:k-MRS-informed-decoder}}: By \eqref{eq:k-MRS_equiv_form}, using the result of Theorem \ref{th:k-MRS-informed-alternative},
\begin{align}
 { R}_{m}^{I}(\Delta )  = \underset{P_{U}, \ \Delta_{u}: \atop \sum \limits_{u} P_{U}(u) \Delta_{u} \leq \Delta } \min \sum \limits_{u \in \cU } P_{U}(u) {\tilde R}(\Delta_{u}), \  \ \ \ \Delta_{\min} \leq \Delta \leq \Delta_{\max}  \label{eq:k-MRS_equiv_form_R_tilde}
\end{align}
where
\begin{align}
 {\tilde R}(\Delta_{u}) = \underset{ P_{X_{\cM} } \delta_{h_{u}(\cdot)} P_{Y_{\cM} | S X_{S}} \atop \mathbbm{E}[d(X_{\cM}, Y_{\cM}) ] \leq \Delta_{u}  }  \min I(X_{S} \wedge Y_{\cM} | S), \ \ \ \Delta_{\min} \leq \Delta_{u} \leq \Delta_{\max} \label{eq:k-MRS-informed-sum-component}
\end{align}
with the pmf $P_{X_{\cM} } \delta_{h_{u}(\cdot)} P_{Y_{\cM} | S X_{S}}$ being understood as $P_{X_{\cM} } \delta_{h(\cdot, u)}  P_{Y_{\cM} | S X_{S}, U = u}$. To simplify notation, the conditioning on $U=u$ will be suppressed except when needed. It suffices to show the existence of a code of rate $ \cong {\tilde R}(\Delta_{u}) $ with distortion $\mathbbm{E}[d(X_{\cM}, Y_{\cM}) ] \ \substack{\sim \\ \leq \\\ } \ \Delta_{u}$.   A concatenation of such codes indexed by $u \in \cU$ yields, in effect, suitable time-sharing among them, leading to the achievability of \eqref{eq:k-MRS_equiv_form_R_tilde}. By Theorem \ref{th:k-MRS-informed-alternative}, in view of the optimality of point-mass samplers, concatenating fixed-set sampling rate distortion codes for conditional sources $P_{X_{\cM}| S = A}, \ A \in \cAk$, will suffice.

\vspace*{0.2cm}

Given any $ \Delta_{\min} \leq \Delta_{u} \leq \Delta_{\max}$, for the minimizer in \eqref{eq:k-MRS-informed-sum-component}, consider the corresponding
\begin{align}
 P_{S|X_{\cM}} = \delta_{h_{u}(\cdot)}, \ \ \Delta_{A_{i}} \triangleq \mathbbm{E}[d(X_{\cM}, Y_{\cM}) | S = A_{i}]  \text{ and }   I(X_{A_{i}} \wedge Y_{\cM} | S = A_{i}), \ \ i \in {\cal M}_{k}.
\end{align}
The associated $\{ (S_{t}, X_{S_{t}}) \}_{t=1}^{\infty}$ is an i.i.d. sequence (cf. Remark (ii) following Definition \ref{d:RDF}). 
The sampling sets characterized   by the conditional point-mass sampler above and the DMMS realizations $x_{\cM}^{n}$, are denoted as $s^{n} (x_{\cM}^{n}) \triangleq \big ( s(x_{\cM 1}), \ldots, s(x_{\cM n})   \big ) $, and hence $S^{n} = s^{n}(X_{\cM}^{n})$.

\vspace*{0.2cm}

The idea behind the remainder of the proof below for each $U=u$ is the following. We collect all those time instants at which a particular $A_{i}$ in $\cAk$ is sampled, with the objective of applying a fixed-set sampling rate distortion code. Since the size of this time-set will vary according to $x_{\cM}^{n} $ in $\cXm^{n}$, the rate of such a code, too, will vary accordingly. However, since we seek fixed rate codes (rather than codes with a desired average rate), we apply fixed-set  sampling codes to subsets of predetermined lengths from among typical sampling sequences in $\cAk^{n}.$

\vspace*{0.2cm}

Fix $\ep >0$ and $0 < \ep' < \ep.$ Ordering the elements of $\cAk$ as in the proof of Theorem \ref{th:k-IRS-informed-decoder}, 
for $n \geq 1$, the sets $\tau_{s^{n}}(A_{i}) \triangleq \{ t:  1 \leq t \leq n, \  s_{t} = A_{i} \}, \ i \in {\cal M}_{k}, $ cover $\{1, \ldots,n \}$; denote the set of the first $\max \left \{ \lceil n( P_{S}(A_{i}) - \ep' ) \rceil ,0 \right \}$ time instants in $\tau_{s^{n}}(A_{i})$ by $ \nu_{A_{i}}$. For the (typical) set  
\begin{align}
{\cal T}_{\ep'}^{(n)} \triangleq  \left \{ s^{n} \in \cAk^{n} : \ \left |  \dfrac{| \tau_{s^{n}} (A_{i})|}{n} - P_{S}(A_{i}) \right | \leq \ep', \ i \in {\cal M}_{k}  \right \},
\end{align}
$P \left (S^{n} \in {\cal T}_{\ep'}^{(n)} \right) \geq 1 - \frac{\ep'}{2}$ for all $n \geq N_{1}(\ep')$, say. 
\vspace*{0.2cm}

Along the lines of proof of Theorem \ref{th:k-IRS-informed-decoder}, for each DMMS with (conditional) pmf $P_{X_{\cM}|S = A_{i}}, \ i \in {\cal M}_{k} $, there exists a code $(f_{A_{i}}, \varphi_{A_{i}}),  \ f_{A_{i}} : {\cal X}_{A_{i}}^{\nu_{A_{i}}} \rightarrow \{1, \ldots, J_{A_{i}} \} $ and $\varphi_{A_{i}}: \{ 1, \ldots, J_{A_{i}} \} \rightarrow \cYm^{\nu_{A_{i}}} $ of rate $\frac{1}{|\nu_{A_{i}}|} \log J_{A_{i}} \leq I(X_{A_{i}} \wedge Y_{\cM} | S = A_{i}) + \frac{\ep'}{2}   $ and with 
\begin{align}
\mathbbm{E} \left [ d \Big ( X_{\cal M}^{\nu_{A_{i}}}, \varphi_{A_{i}} \big ( f_{A_{i}}(X_{A_{i}}^{\nu_{A_{i}}}) \big ) \Big ) \Big | S^{\nu_{A_{i}}} = A_{i}^{\nu_{A_{i}}} \right ] \leq \Delta_{A_{i}} + \frac{\ep'}{2}   
\end{align}
for all $| \nu_{A_{i}} | \geq N_{A_{i}} \left (\frac{\ep'}{2} \right ). $ 

\vspace*{0.2cm}

A (composite) code $(f, \varphi_{S})$, with $f$ taking values in $\cJ \triangleq \mathop{\mbox{\large $\times$}} \limits_{i=1}^{M_{k}} \left \{ 1, \ldots, J_{A_{i}} \right \} $   is constructed as follows. The encoder $f$ consists of a concatenation of encoders defined by
\begin{align}
 f \left ( s^{n}(x_{\cM}^{n})  ; x_{s_{1}}, \ldots, x_{s_{n}} \right ) = \begin{cases}
                                   \left (  f_{A_{1}} \left ( x_{A_{1}}^{ \nu_{A_{1}} }  \right ),  \ldots, f_{A_{M_{k}}} \left ( x_{A_{M_{k}}}^{ \nu_{A_{M_{k}}} }  \right ) \right), \ \ \ &s^{n}(x_{\cM}^{n}) \in {\cal T}_{\ep'}^{(n)},  \  x_{\cM}^{n} \in \cXm^{n} \\
                                   (1, \ldots,1), \ \ \ & s^{n}(x_{\cM}^{n}) \notin {\cal T}_{\ep'}^{(n)}, \ x_{\cM}^{n} \in \cXm^{n}.
                                  \end{cases} 
\end{align}

\noindent For $t = 1, \ldots,n,$ and $(j_{1}, \ldots, j_{M_{k}}) \in \cJ,$ the informed decoder $\varphi_{S}$ is given by
\begin{align}
 \Big ( \varphi_{S} \big ( s^{n} , (j_{1}, \ldots, j_{M_{k}}) \big) \Big )_{t} = \begin{cases}
                                                                     \left (\varphi_{A_{i}}  \left (x_{A_{i}}^{ \nu_{A_{i}}} \right) \right )_{t} ,  \ & s^{n} \in {\cal T}_{\ep'}^{(n)} \text{ and } t \in \nu_{A_{i}}, \  i \in {\cal M}_{k},   \\
                                                                   y_{1}, \ \ \ & \text{otherwise,}
                                                                  \end{cases}
\end{align}
where $y_{1}$ is a fixed but arbitrary symbol in $\cYm.$

\vspace*{0.2cm}

The rate of the code is  
\begin{align}
\dfrac{1}{n} \log || \cJ || & = \dfrac{1}{n} \sum \limits_{ i = 1}^{M_{k}} \log J_{A_{i}} \\
& \leq  \dfrac{1}{n} \left ( \sum \limits_{ i = 1}^{M_{k}}  |\nu_{A}| \left ( I(X_{A_{i}} \wedge Y_{\cM} | S = A_{i}  ) + \frac{\ep'}{2} \right)  \right ) \\
& \leq   \sum \limits_{ i = 1}^{M_{k}}  P_{S}(A_{i}) \left ( I(X_{A_{i}} \wedge Y_{\cM} | S = A_{i}  ) + \frac{\ep'}{2} \right )     \\
& =  \sum \limits_{ i = 1}^{M_{k}}  P_{S}(A_{i}) I \left (X_{A_{i}} \wedge Y_{\cM} | S = A_{i}  \right )  + \frac{\ep'}{2} \leq I(X_{S} \wedge Y_{\cM} | S ) + \ep \label{eq:thm2-ach-eq1}.
\end{align}

\vspace*{0.2cm}

Defining $d_{\max} \triangleq \underset{ (x_{\cM}, y_{\cM}) \in \cXm \times \cYm } \max d(x_{\cM} ,y_{\cM}) $, and with $Y_{\cM}^{n}$ denoting the output of the decoder, we have
\begin{align}
 \mathbbm{E}[ d(X_{\cM}^{n},Y_{\cM}^{n}) ] & =  \ \mathbbm{E} \left [  \mathbbm{E} \left [ d(X_{\cM}^{n}, Y_{\cM}^{n}) \big | S^{n} \right ] \right ] \\ 
 & = \ \sum_{s^{n} \in {\cal T}_{\ep'}^{(n)}} P_{S^{n} }(s^{n})  \mathbbm{E} \big [ d(X_{\cM}^{n}, Y_{\cM}^{n}) \big | S^{n} = s^{n} \big ] \  +  \sum_{s^{n} \notin {\cal T}_{\ep'}^{(n)}} P_{S^{n}}(s^{n}) \mathbbm{E} \left [ d(X_{\cM}^{n}, Y_{\cM}^{n}) \big | S^{n} = s^{n} \right ]  \\
& \leq \ \sum_{s^{n} \in {\cal T}_{\ep'}^{(n)}} P_{S^{n} }(s^{n}) \sum_{ i = 1 }^{M_{k}} \frac{|\nu_{A_{i}}| }{n} \mathbbm{E} \left [ d \Big ( X_{\cal M}^{\nu_{A_{i}}}, \varphi_{A_{i}} \big ( f_{A_{i}}(X_{A_{i}}^{\nu_{A_{i}}}) \big ) \Big ) \Big | S^{\nu_{A_{i}}} = A_{i}^{\nu_{A_{i}}} \right ] \ \\ & \ \ \ \  + \ \frac{1}{n} \sum_{s^{n} \in {\cal T}_{\ep'}^{(n)}} P_{S^{n} }(s^{n})  \sum \limits_{i = 1}^{M_{k}} \sum \limits_{t \in \tau_{s^{n}}(A_{i}) \setminus \nu_{A_{i}}}  \mathbbm{E} \left [d(X_{\cM}, Y_{\cM t}) | S^{n} = s^{n} \right ]  \  +  \sum_{s^{n} \notin {\cal T}_{\ep'}^{(n)}} P_{S^{n}}(s^{n}) { d}_{\max} \\
& \leq \  \sum_{ i = 1 }^{M_{k}}  P_{S}(A_{i}) \left ( \Delta_{A_{i}} + \frac{\ep'}{2} \right )   \  + \ \left ( 1 - \frac{\sum \limits_{i=1}^{M_{k}} |\nu_{A_{i}}|  }{n}  \right ) d_{\max}  \  +   \frac{\ep'}{2} d_{\max} \\
& \leq \    \Delta_{ {u}} + \frac{\ep'}{2}     \  + \ \left ( 2 M_{k} {\ep'}  \right ) d_{\max}  \  +   \frac{\ep'}{2} d_{\max} \\
& < \Delta_u +  \epsilon \label{eq:thm2-ach-eq2}
\end{align}
for all $n$ large enough. The proof is completed by noting that for $n$ large enough \eqref{eq:thm2-ach-eq1} and \eqref{eq:thm2-ach-eq2} hold simultaneously and time-sharing between the codes corresponding to $U=u, \ u \in \cU,$ completes the proof.

\qed



\vspace*{0.3cm}

\noindent {\bf  Theorem \ref{th:k-MRS-uninformed-decoder}:} The proof is similar to that of Proposition \ref{th:k-FS-RD} with the i.i.d. sequence $\{ X_{A {t}} \}_{t=1}^{\infty}$ replaced by the i.i.d. sequence $\{ S_{t}, X_{S_{t}} \}_{t=1}^{\infty}$ with joint pmf $P_{S X_{S}}$ obtained from \eqref{eq:k-MRS-RD-uninformed} and a modified distortion measure ${\tilde d} \left( (s,x_{s}), y_{\cM} \right) \triangleq \mathbbm{E} \big [d(X_{\cM},y_{\cM}) | S = s, X_{S} = x_{s} \big ] $. The details, identical to those in the achievability proof of Proposition \ref{th:k-FS-RD}, are omitted.

\qed




\vspace*{0.5cm  }
\subsection{Converse proofs}

\noindent Separate converse proofs can be provided for Proposition \ref{th:k-FS-RD} and Proposition \ref{prop:SMRS_SIRS}. However, in order to highlight the underlying ideas economically, we develop the proofs in a unified manner. Specifically, in contrast with the achievability proofs above, our converse proofs are presented in the order of weakening power of the sampler, viz., $k$-MRS, $k$-IRS and fixed-set sampler. We begin with the proof of Lemma \ref{l:SRDf_convexity} followed by pertinent technical results before turning to Proposition \ref{th:k-FS-RD} and Proposition \ref{prop:SMRS_SIRS}.

\vspace*{0.2cm}


\noindent {\bf  Lemma \ref{l:SRDf_convexity}}: We need to prove only that the right-sides of \eqref{eq:k-FS-RD}, \eqref{eq:k-IRS-informed-PMF} and \eqref{eq:k-MRS-RD-informed} are convex and continuous, since they are evidently finite-valued and nonincreasing in $\Delta.$ The convexity of the right-side of \eqref{eq:k-FS-RD} on $[\Delta_{\min, A}, \Delta_{\max}]$ is a standard consequence of the convexity of $I(X_{A} \wedge Y_{\cM}) = I \left (P_{X_{A}}, P_{Y_{\cM} | X_{A}} \right)$ in $P_{Y_{\cM}|X_{A}}$ and the convexity of the constraint set in \eqref{eq:k-FS-RD}. The convexity of the right-sides of \eqref{eq:k-IRS-informed-PMF} and \eqref{eq:k-MRS-RD-informed} is immediate by the remarks following Proposition \ref{prop:k-IRS-equiv} and Theorem \ref{th:k-MRS-informed-decoder}, and their continuity for $\Delta > \Delta_{\min}$ is a consequence. Continuity at $\Delta = \Delta_{\min}$ in \eqref{eq:k-FS-RD}, \eqref{eq:k-IRS-informed-PMF} and \eqref{eq:k-MRS-RD-informed} holds, for instance, as in (\cite{CsiKor11}, Lemma 7.2). \qed


\noindent 
\begin{lemma}
\label{l:iid-converse}
Let the finite-valued rvs $A^{n},B^{n},C^{n},D^{n}$ be such that $(A_{t}, B_{t}), \ t = 1, \ldots,n,$ are mutually independent and satisfy
\begin{align}
\label{eq:Block_markov_prop}
 B^{n} \MC A^{n}, C^{n} \MC D^{n}
 \end{align}
 and
\begin{align}
\label{eq:Sampler_Markov_given}
 C_{t} \MC A_{t}, B_{t}, C^{t-1} \MC A^{n\setminus t}, B^{n \setminus t} , \ \ \ \ t=1, \ldots,n,
\end{align}
where $A^{n\setminus t} = A^{n} \setminus A_{t}.$
Then, the following hold for $t=1, \ldots,n$:
\begin{align}
\label{eq:ABCD_lemma_claim1}
 I( A^{t}, B^{t}, C^{t} \wedge A_{t+1}^{n}, B_{t+1}^{n} ) = 0 \ ;
\end{align}
\begin{align}
\label{eq:ABCD_lemma_claim2}
 A_{t}, B_{t}  \MC C^{t} \MC A^{n \setminus t}, B^{n \setminus t}, C_{t+1}^{n} \ \ ;
\end{align}
and
\begin{align}
 \label{eq:ABCD_lemma_claim3}
B_{t} \MC A_{t}, C^{t} \MC D_{t}.
\end{align}
  
\end{lemma}

\noindent {\bf Proof:} First, \eqref{eq:ABCD_lemma_claim1} is true by the following simple observation: for $t = 1 , \ldots, n,$
\begin{align}
 I( A^{t}, B^{t}, C^{t} \wedge A_{t+1}^{n}, B_{t+1}^{n} ) & = I( A^{t}, B^{t} \wedge A_{t+1}^{n}, B_{t+1}^{n} ) + I( C^{t} \wedge A_{t+1}^{n}, B_{t+1}^{n} | A^{t}, B^{t}) = 0 \label{eq:gen_ABCD_sampler_eq1} 
\end{align}
where the first term in the sum above is zero by the mutual independence of $ (A_{t}, B_{t} ), \ t = 1, \ldots,n,$ and the second term equals zero by \eqref{eq:Sampler_Markov_given}. 
Next, the claim \eqref{eq:ABCD_lemma_claim1} and the Markov property \eqref{eq:Sampler_Markov_given} imply that for $t = 1, \ldots,n,$
\begin{align}
\label{eq:ABCD_lemma_eq1}
 I(A_{t}, B_{t}, C_{t} \wedge A^{t-1}, B^{t-1} | C^{t-1}) = I(A_{t}, B_{t} \wedge A^{t-1}, B^{t-1} | C^{t-1})  + I(C_{t} \wedge A^{t-1}, B^{t-1} | A_{t}, B_{t}, C^{t-1})= 0.
\end{align}

\noindent The claim \eqref{eq:ABCD_lemma_claim2} now follows, since
\begin{align}
 I(A_{t}, B_{t} \wedge A^{n \setminus t}, B^{n \setminus t}, C_{t+1}^{n} | C^{t})  & =  I(A_{t}, B_{t} \wedge A^{t-1}, B^{t-1} | C^{t}) + I(A_{t}, B_{t} \wedge A_{t+1}^{n}, B_{t+1}^{n} | A^{t-1}, B^{t-1}, C^{t})  \\ 
 &\hspace*{0.3cm} \  +  I(A_{t}, B_{t} \wedge C_{t+1}^{n} | A^{n \setminus t}, B^{ n \setminus t}, C^{t}) \\
 & = 0
\end{align}
where the first term in the sum above is zero by \eqref{eq:ABCD_lemma_eq1}, and the latter two terms are zero by \eqref{eq:gen_ABCD_sampler_eq1} and \eqref{eq:Sampler_Markov_given}, respectively.  

\vspace*{0.2cm}

Now using \eqref{eq:Block_markov_prop},
\begin{align}
 0 & = I(B^{n} \wedge D_{t} |A^{n}, C^{n})  \\
   & = \sum \limits_{t=1}^{n} I(B_{t} \wedge D_{t} | B^{t-1},A^{n}, C^{n} )  \\
   & = \sum \limits_{t=1}^{n} I(B_{t} \wedge D_{t} | B^{t-1}, A_{t}, A^{n \setminus t}, C^{t}, C_{t+1}^{n} )  \\
   & = \sum \limits_{t=1}^{n} \left [ I(B_{t} \wedge B^{t-1}, A^{n \setminus t}, C_{t+1}^{n},D_{t} | A_{t}, C^{t} ) -   I(B_{t} \wedge B^{t-1},A^{n \setminus t}, C_{t+1}^{n}| A_{t},C^{t} )  \right ] \\
   & = \sum \limits_{t=1}^{n} I( B_{t} \wedge B^{t-1}, A^{n \setminus t},C_{t+1}^{n}, D_{t} | A_{t}, C^{t} ) \ \ \ \text{by }   \eqref{eq:ABCD_lemma_claim2} \\
   & \geq \sum_{t=1}^{n} I(B_{t} \wedge D_{t} |A_{t}, C^{t}),
\end{align} 
so that the claim \eqref{eq:ABCD_lemma_claim3} follows.
\qed

\vspace*{0.5cm}

We now prove Proposition \ref{prop:SMRS_SIRS} which, in effect, implies the converse proofs for Theorem \ref{th:k-MRS-informed-decoder}, Theorem \ref{th:k-IRS-informed-decoder} and   Proposition \ref{th:k-FS-RD}. Specifically, a converse is fashioned for $R_{ms}^{I}(\Delta),$ with those for $R_{is}(\Delta)$ and $R_{A}(\Delta)$ emerging along the way.
 
\vspace*{0.2cm}

Let $ \left ( \{ P_{S_{t}| X_{\cM t} S^{t-1}  } \}_{t=1}^{n},  f, \varphi_{S} \right )$ be an $n$-length strong $k$-MRS block code with decoder output \\
$Y_{\cM}^{n} = \varphi_{S} \big( S^{n}, f(S^{n},X_{S}^{n} ) \big)$ and satisfying $\mathbbm{E} \left [ d \big( X_{\cM}^{n}, Y_{\cM}^{n} \big) \right] \leq \Delta$. The hypothesis of Lemma \ref{l:iid-converse} with $A^{n} = X_{S}^{n}, \ B^{n} = X_{S^{c}}^{n}, \ C^{n} = S^{n} \ \text{and} \ D^{n} = Y_{\cM}^{n}$ is met since 
\begin{align}
 X_{S^{c}}^{n} \MC S^{n}, X_{S}^{n} \MC Y_{\cM}^{n}
\end{align}
and by \eqref{eq:k_SMRS_def},
\begin{align}
 P_{S_{t} | X_{\cM}^{n} S^{t-1}} = P_{S_{t} | X_{\cM t} S^{t-1}}.
\end{align}
Then by Lemma \ref{l:iid-converse}, for $t=1, \ldots,n$,
\begin{align}
\label{eq:converse_causal_sampler_prop}
 I(S^{t-1} \wedge X_{\cM t}) = 0,
\end{align}
\begin{align}
\label{eq:Sampler_markov_converse}
 X_{\cM t} \MC S^{t} \MC X_{\cM}^{n \setminus t}, X_{S }^{n \setminus t}, S_{t+1}^{n}
\end{align}
and
\begin{align}
\label{eq:converse_code_markov}
 X_{S_{t}^{c}} \MC S^{t}, X_{S_{t}} \MC Y_{\cM t}.
\end{align}

\noindent Denoting by $||f ||$ the cardinality of the range space of the encoder $f$, the rate $R$ of the code satisfies 
\begin{align}
 nR &  = \log ||f||  \geq     H \big (  f (S^{n}, X_{S}^{n} ) \big )   \\ 
   & \geq H \Big (\varphi_{S} \big( S^{n}, f (S^{n}, X_{S}^{n} ) \big ) | S^{n} \Big)  =  H \big (Y_{\cM}^{n} | S^{n} \big) \\
   & =  H \big (Y_{\cM}^{n} | S^{n} \big) -  H\big (Y_{\cM}^{n} | S^{n}, X_{S}^{n} \big)   \\
   & = I \big( X_{S}^{n} \wedge Y_{\cM}^{n} | S^{n} \big)  \\
   & = \sum_{t=1}^{n} \left( H( X_{S_{t}} | S^{t}, S_{t+1}^{n},  X_{S }^{t-1} ) - H( X_{S_{t}} | S^{t}, S_{t+1}^{n}, X_{S }^{t-1} , Y_{\cM}^{n} ) \right)   \\
   & \geq \sum_{t=1}^{n} \big( H( X_{S_{t}} |S^{t} ) - H( X_{S_{t}} | S^{t}, Y_{\cM t} ) \big) \label{eq:gen_converse_eq1}   \\
   & = \sum_{t=1}^{n} I(X_{S_{t}} \wedge Y_{\cM t} | S^{t})    \label{eq:converse_generic_eq2}
\end{align}
where \eqref{eq:gen_converse_eq1} follows from \eqref{eq:Sampler_markov_converse}. Denote $\mathbbm{E}[d( X_{\cM t}, Y_{\cM t} )]$ by $\Delta_{t}$.

\vspace*{0.2cm}

For the strong $k$-MRS code above, in \eqref{eq:converse_generic_eq2} using \eqref{eq:converse_causal_sampler_prop} and \eqref{eq:converse_code_markov}, we get 
\begin{align}
 I(X_{S_{t}} \wedge Y_{\cM t} | S^{t}) & \geq \ \underset{  P_{S^{t-1}} P_{X_{\cM t}}  P_{S_{t}| X_{\cM t} S^{t-1} }  P_{Y_{\cM}| S_{t} X_{S_{t}} S^{t-1} } \atop \mathbbm{E}[d(X_{\cM t}, Y_{\cM t})  ] = \Delta_{t} } \min I(X_{S_{t}} \wedge Y_{\cM t} | S_{t}, S^{t-1} ) \label{eq:converse_generic_eq3} \\
 & \geq \ \underset{  P_{U_{t}} P_{X_{\cM t}}  P_{S_{t}| X_{\cM t} U_{t} }  P_{Y_{\cM}| S_{t} X_{S_{t}} U_{t} } \atop \mathbbm{E}[d(X_{\cM t}, Y_{\cM t})  ] \leq \Delta_{t} } \min I(X_{S_{t}} \wedge Y_{\cM t} | S_{t}, U_{t} ), \label{eq:converse_generic_eq4} 
\end{align}
where  $U_{t}$ is a rv taking values in a set of cardinality $|\cAk|^{t-1}.$
The existence of the minima in \eqref{eq:converse_generic_eq3} and \eqref{eq:converse_generic_eq4} comes from the continuity of the conditional mutual information terms over compact sets of pmfs.

\vspace*{0.2cm}

By the Carath\'eodory theorem \cite{CoverThomas}, every point in the convex hull of the set ${\cal C} = \Big \{ \big( \mathbbm{E}[d(X_{\cM}, Y_{\cM})], I(X_{S} \wedge Y_{\cM} | S) \big ) : X_{\cM} \MC S,X_{S} \MC Y_{\cM} \Big \} \subset \mathbbm{R}^{2} $ can be represented as a convex combination of at most three points in ${\cal C}.$ Hence, to describe every element in the set 
\begin{align}
\Big \{ \big( \mathbbm{E}[d(X_{\cM t}, Y_{\cM t})], I(X_{S_{t} } \wedge Y_{\cM t} | S_{t}, U_{t}) \big )  : P_{U_{t} X_{\cM t} S_{t} Y_{\cM t} } =  P_{U_{t}} P_{X_{\cM t}} P_{S_{t}|X_{\cM t} U_{t}} P_{Y_{\cM t}|S_{t} X_{S_{t}} U_{t}} \Big \}, 
\end{align}
it suffices to consider a rv $U_{t}$ with support of size three. (For $t=1$, this assertion is straightforward.) Consequently, the right-side of \eqref{eq:converse_generic_eq4} equals $R_{m}^{I}(\Delta_{t})$ (cf. \eqref{eq:k-MRS-RD-informed}). Using the convexity of $R_{m}^{I}(\Delta)$ in $\Delta$, we get from \eqref{eq:converse_generic_eq2} that
\begin{align}
 nR & \geq \sum \limits_{t=1}^{n} R_{m}^{I}(\Delta_{t}) \\
 & \geq n R_{m}^{I} \left ( \frac{1}{n} \sum \limits_{t=1}^{n} \Delta_{t} \right  ) \label{eq:converse_generic_eqtest3} \\
 & \geq n R_{m}^{I}( \Delta),
\end{align}
i.e., $R \geq R_{m}^{I}(\Delta), \ \Delta \geq \Delta_{\min},$ thereby completing the converse proof for a strong $k$-MRS and Theorem \ref{th:k-MRS-informed-decoder}.

\vspace*{0.2cm}

Next, an $n$-length strong $k$-IRS code and fixed-set sampler code can be viewed as restrictions of the strong $k$-MRS code above. Specifically, the strong $k$-IRS and fixed-set sampler respectively entail replacing $P_{S_{t}|X_{\cM t} S^{t-1}}$ by $P_{S_{t}|S^{t-1}}$ and $P_{S_{t}} = \mathbbm{1}(S_{t} = A)$. Counterparts of \eqref{eq:converse_generic_eq3} and \eqref{eq:converse_generic_eq4} hold with the mentioned replacements. For a strong $k$-IRS, upon replacing $P_{S_{t}|X_{\cM t} S^{t-1}}$ with $P_{S_{t}|S^{t-1}}$, we observe that the right-side of \eqref{eq:converse_generic_eq4}, viz.
\begin{align}
 \underset{P_{U_{t}} P_{X_{\cM t}} P_{S_{t}|U_{t}} P_{Y_{\cM t}|S_{t} X_{S_{t}} U_{t}} \atop \mathbbm{E}[d(X_{\cM t}, Y_{\cM t})] \leq \Delta_{t} } \min I(X_{S_{t}} \wedge Y_{\cM t} | S_{t}, U_{t}) 
\end{align}
is now the lower convex envelope of the SRDf for a $k$-IRS, already convex in distortion, and hence, equals $R_{i}(\Delta_{t})$ itself. Thus, \eqref{eq:converse_generic_eq4} becomes
\begin{align}
 I(X_{S_{t}} \wedge Y_{\cM t} | S^{t}) \geq \underset{  P_{X_{\cM t} } P_{S_{t}}  P_{Y_{\cM t }| S_{t} X_{S_{t}} } \atop \mathbbm{E}[d(X_{\cM t}, Y_{\cM t}) ] \leq \Delta_{t} } \min I(X_{S_{t}} \wedge Y_{\cM t} | S_{t}) = R_{i}( \Delta_{t}). \label{eq:converse_generic_eq5}
\end{align}
Combining \eqref{eq:converse_generic_eq2} and \eqref{eq:converse_generic_eq5}, we get along the lines of \eqref{eq:converse_generic_eqtest3} that $R \geq R_{i} (\Delta), \ \Delta \geq \Delta_{\min},$ which gives the converse proof for a strong $k$-IRS and Theorem \ref{th:k-IRS-informed-decoder}.

\vspace*{0.2cm}

In a manner analogous to a strong $k$-IRS, for a fixed-set sampler the convexity of $R_{A}(\Delta)$ in $\Delta$ implies that the counterpart of the right-side of \eqref{eq:converse_generic_eq4}, with $P_{S_{t}|X_{\cM t} U_{t}} $ replaced by $\mathbbm{1}(S_{t} = A)$, simplifies to $R_{A}(\Delta_{t})$. As in \eqref{eq:converse_generic_eqtest3}, it follows that $R \geq R_{A}(\Delta), \ \Delta \geq \Delta_{\min, A},$ which gives the converse for Proposition \ref{th:k-FS-RD}.

\qed

\section{Conclusion}
Our new framework of sampling rate distortion describes the centralized sampling of fixed-size subsets of the components of a DMMS, followed by encoding and  lossy reconstruction of the full DMMS. Specifically, we examine the tradeoffs between sampling strategy, optimal encoding rate and distortion in reconstruction as characterized by a sampling rate distortion function. Three sampling strategies are considered: fixed-set sampling, independent random sampling and memoryless random sampling; in the latter two settings, the decoder may or may not be informed of the sampling sequence.

\vspace*{0.2cm}
Single-letter characterizations of the SRDf are provided for the sampling strategies above but for a memoryless random sampler with uninformed decoder. In the last case, an achievability proof yields an upper bound for the SRDf whose tightness is unknown. This upper bound in Theorem \ref{th:k-MRS-uninformed-decoder} can be convexified by means of a time-sharing random variable whereupon the modified bound becomes tight. However, it remains open whether such time-sharing is necessary for convexification.


\section{Acknowledgement}
The authors thank Himanshu Tyagi for many helpful discussions and for suggesting the model with strong sampler.


\renewcommand\baselinestretch{0.9}
{\small
\providecommand{\bysame}{\leavevmode\hbox to3em{\hrulefill}\thinspace}

}

\appendix
\label{s:Appendix}
\section*{Proof of \eqref{eq:Alternative_ach_eq4} }

We show that the minimum of \eqref{eq:Alternative_ach_eq2} with respect to $\left ( P_{U}, P_{S|X_{\cM} U }, P_{Y_{\cM}|S X_{S} U}  \right)$ is attained by $P_{S|X_{\cM} U}$ of the form $\delta_{h (\cdot) }$. The Lagrangian is
\begin{align}
 G_{q}(\lambda) & =   \underset{ P_{U}, P_{S|X_{\cM}U}, \atop  P_{ Y_{\cM} | S X_{S}U }, Q_{ Y_{\cM} |  S U } } \min \   D \left (  P_{Y_{\cM}|SX_{S} U} \big | \big | Q_{Y_{\cM}|  S U} \big | P_{S X_{S} U} \right )  +  \lambda \mathbbm{E} \big [ d(X_{\cM}, Y_{\cM})  \big ]    \label{eq:Topsoe_usage} \\
 & = \underset{ P_{U}, Q_{Y_{\cM}| S U}, \atop  P_{ Y_{\cM} | S X_{S}U } } \min \   \sum \limits_{u,x_{\cM}} P_{U X_{\cM}}(u, x_{\cM})  \underset{ P_{S|X_{\cM}U}   } \min \sum \limits_{s \in \cAk} P_{S|  X_{\cM} U}(s|x_{\cM},u)  
 \Bigg ( \mathbbm{E} \Big [ \log \frac{ P_{Y_{\cM}|S X_{S} U} (Y_{\cM} | s,x_{s},u ) }{Q_{Y_{\cM}|SU} (Y_{\cM}| s, u) }  \\ & \hspace*{10.7 cm} +  \lambda d(x_{\cM}, Y_{\cM} ) \Big | S = s, X_{S} = x_{s}, U = u \Big ] \Bigg )   
 \end{align}
where the expectation above is with respect to $P_{Y_{\cM}|S= s , X_{S} = x_{s}, U = u}.$ Noting that the term in $\Bigg( \cdots \Bigg ) $ is a function of $s,x_{\cM},u,$ we get 
 \begin{align}
 G_{q}(\lambda) & = \underset{ P_{U}, Q_{Y_{\cM}| S U}, \atop   P_{ Y_{\cM} | S X_{S}U } } \min \   \sum \limits_{u,x_{\cM}} P_{U X_{\cM}}(u, x_{\cM})  \ \underset{ s \in \cAk } \min \  
 \Bigg ( \mathbbm{E} \Big [ \log \frac{ P_{Y_{\cM}|S X_{S} U} (Y_{\cM} | s,x_{s},u ) }{Q_{Y_{\cM}|SU} (Y_{\cM}| s, u) }  \\ & \hspace*{7.1 cm} +  \lambda d(x_{\cM}, Y_{\cM} ) \Big | S = s, X_{S} = x_{s}, U = u \Big ] \Bigg )  \\
  & =   \underset{ P_{U},   Q_{Y_{\cM}|   S U}, \atop   P_{ Y_{\cM} | S X_{S}U } } {\min} \   \sum \limits_{u,x_{\cM}} P_{U X_{\cM}}(u, x_{\cM}) \ \underset{ \delta_{h( \cdot )} } \min  \sum \limits_{s \in \cAk} \delta_{h(x_{\cM}, u) }(s)  \Bigg ( \mathbbm{E} \Big [ \log \frac{ P_{Y_{\cM}|S X_{S} U} (Y_{\cM} | s,x_{s},u ) }{Q_{Y_{\cM}|SU} (Y_{\cM}| s, u) }  \\ & \hspace*{9cm} + \lambda d(x_{\cM}, Y_{\cM} ) \Big | S = s, X_{S} = x_{s}, U = u \Big ] \Bigg )    \\
    & = \underset{ P_{U},  \delta_{ h( \cdot ) },  Q_{ Y_{\cM} |  S U } , \atop  P_{ Y_{\cM} | S X_{S}U } } \min \    D \left (  P_{Y_{\cM}|SX_{S} U} \big | \big | Q_{Y_{\cM}|  S U} \big | P_{S X_{S} U} \right )  +  \lambda \mathbbm{E} \big [ d(X_{\cM}, Y_{\cM})  \big ]    \\   
 & = G_{r}( \lambda). \label{eq:Alternative_ach_eq3}
\end{align}


\begin{thebibliography}{20}


\bibitem{Ber71}
T.~Berger,
{\it Rate Distortion Theory: A Mathematical Basis for
Data Compression}, Prentice-Hall, Englewood Cliffs, NJ, 1971.


\bibitem{Ber78}
T.~Berger, 
``Multiterminal source coding,'' in {\it The Information Theory
Approach to Communications}, G. Longo, Ed. Vienna/New York:
Springer-Verlag, 1978, vol. 229, CISM Courses and Lectures, pp.
171--231.


\bibitem{BodaNarayan14}
V.~P.~Boda and P.~Narayan, ``Sampling rate distortion,''  {\it Proceedings of the IEEE International Symposium on Information Theory (ISIT), 2014}, pp. 3057--3061, June 29-July 4 2014.



\bibitem{BodaNarayan15}
V.~P.~Boda and P.~Narayan, ``Memoryless sampling rate distortion,''  {\it 53rd Annual Allerton Conference on Communication, Control, and Computing, 2015}, pp. 919-923, Sept. 30-Oct. 2 2015.


\bibitem{BodaNarayan16}
V.~P.~Boda and P.~Narayan, ``Independent and memoryless sampling rate distortion,''  {\it Proceedings of the IEEE International Symposium on Information Theory (ISIT), 2016}, pp.2968-2972, July 10-15 2016.


\bibitem{CanRomTao06}
E.~J. Cand\`es, J.~Romberg and T.~Tao, 
``Robust uncertainty principles: Exact signal reconstruction
from highly incomplete frequency information,''
 {\it IEEE Trans. Inform. Theory}, 
vol.~52, no.~2, pp.~489--509, Feb. 2006.


\bibitem{CanTao06}
E.~J. Cand\`es and T.~Tao, 
``Near-optimal signal recovery from random
projections: Universal encoding strategies,''
{\it IEEE Trans. Inform. Theory}, 
vol.~52, no.~12, pp.~5406--5425, Dec. 2006.


\bibitem{ChenRanZhanVet13}
Z.~ Chen, J.~Ranieri, R.~Zhang and M.~Vetterli, ''DASS: Distributed Adaptive Sparse Sensing``, {\it IEEE Trans.  Wireless Comm.}, vol.~14, no.~5, pp. 2571--2583, May 2015.

\bibitem{CoverThomas}
T.~M.~Cover and A.~J.~Thomas, {\it Elements of Information Theory}, John Wiley \& Sons, 2012.


\bibitem{CsiKor11}
I.~Csisz\'ar and J.~K\"{o}rner, {\it Information Theory: Coding Theorems for Discrete Memoryless Systems},  Cambridge University Press, 2011.


\bibitem{DobTsy62}
R.~L.~Dobrushin and B.~S.~Tsybakov, ``Information transmission with
additional noise," {\it IRE Trans. Inform. Theory}, vol.~8, no.~5, pp. 293--304, Sept. 1962.


\bibitem{Don06}
D.~L. Donoho, 
``Compressed sensing,'' {\it IEEE Trans. Inform. Theory}, vol.~52, no.~4, pp.~1289--1306, April 2006.



\bibitem{FleRanGoyRam06}
A.~K. Fletcher, S.~Rangan, V.~K. Goyal and K.~Ramachandran,
``Denoising by sparse approximation: Error bounds based on
rate-distortion theory," {\it EURASIP  J. Adv. Sig. Proc.}, vol. 1, pp. 1--19, Dec. 2006.


\bibitem{FleRanGoy07}
A.~K. Fletcher, S.~Rangan and V.~K. Goyal,
 ``On the rate-distortion performance of compressed sensing," 
 {\it Proceedings of IEEE International Conference on Acoustics, Speech and Signal Processing},  vol.~3, pp. III-885--III-888, 15-20 April 2007.
 
 

 \bibitem{HorRiyLuVet10}
A.~Hormati, O.~Roy, Y.~M. Lu and M.~Vetterli,
 ``Distributed sampling of signals linked by sparse filtering: theory and applications,"
 {\it IEEE Trans. Sig. Proc.}, vol.~58, no.~3, pp. 1095--1109, March 2010. 
 
\bibitem{IshKunRam03}
P.~Ishwar, A.~Kumar and K.~Ramachandran,
``On distributed sampling in dense sensor networks: a ``bit conservation principle,"
{\it International Symposium on Information Processing in
Sensor Networks} (IPSN), Palo Alto, CA, April 2003.
 

 
\bibitem{KashLasXiaLiu05}
 A.~Kashyap, L.~A. Lastras-Montano, C.~Xia and L.~Zhen, ``Distributed source coding in dense sensor networks," {\it Proceedings of Data Compression Conference, 2005}, pp. 13--22, 29-31 March 2005.
 
\bibitem{KawDem94}
T.~Kawabata and A.~Dembo,
``The rate-distortion dimension of sets and measures,"
{\it IEEE Trans. Inform. Theory}, vol.~40, no.~5, pp. 1564--1572, Dec. 1994.


\bibitem{KipnisGold16}
A.~Kipnis, A.~J.~Goldsmith, Y.~C.~Eldar and T.~Weissman,
``Distortion rate function of sub-Nyquist sampled Gaussian sources,"
{\it IEEE Trans. Inform. Theory}, vol.~62, no.~1, pp. 401--429, Jan. 2016.


\bibitem{KonTelVet12}
R.~L. Konsbruck, E.~Telatar and M.~Vetterli,
 ``On sampling and coding for distributed acoustic sensing," 
 {\it IEEE Trans. Inform. Theory}, vol.~58, no.~5, pp. 3198--3214, May 2012.
 
\bibitem{LiuSimErk12}
 X.~Liu, O.~Simeone and E.~Erkip, ``Lossy computing of correlated sources with fractional sampling," {\it Proceedings of the IEEE Information Theory Workshop (ITW), 2012},  pp. 232--236, 3-7 Sept. 2012.
 
 
\bibitem{NeuPra11}
D.~L. Neuhoff and S.~S. Pradhan,
 ``Information rates of densely sampled Gaussian data," 
 {\it Proceedings of the IEEE International Symposium on Information Theory Proceedings (ISIT), 2011}, pp. 2776--2780, July 31 2011-Aug. 5 2011.

\bibitem{NeuPra12}
D.~L. Neuhoff and S.~S. Pradhan,
 ``Rate-distortion behavior at low distortion for densely sampled Gaussian data," 
 {\it Proceedings of the IEEE International Symposium on Information Theory Proceedings (ISIT), 2012}, pp. 358--362, 1-6 July 2012.

 \bibitem{PraNeu07}
S.~S. Pradhan and D.~L. Neuhoff,
 ``Transform coding of densely sampled Gaussian data," 
 {\it Proceedings of the IEEE International Symposium on  Information Theory (ISIT), 2007}, pp. 1111--1114, 24-29 June 2007.

 
 \bibitem{RanVet13}
 J.~Ranieri, A.~ Chebira and M.~Vetterli,
 ''Near-Optimal Sensor Placement for Linear Inverse Problems``, {\it IEEE Trans. Sig. Proc.}, vol.~62, no.~5, pp. 1135--1146, March 2014.
   
 
\bibitem{ReeGas12}
G.~Reeves and M.~Gastpar,
 ``The sampling rate-distortion tradeoff for sparsity pattern recovery in compressed sensing," 
 {\it IEEE Trans. Inform. Theory}, vol.~58, no.~5, pp. 3065--3092, May 2012.
 
\bibitem{SunGoy12}
J.~Z. Sun and V.~K. Goyal, ``Intersensor collaboration in distributed quantization networks,'' {\it IEEE Trans.  Comm.}, vol.~61, no.~9, pp. 3931--3942, Sept. 2013.


\bibitem{UnnVet12a}
J.~Unnikrishnan and M.~Vetterli,
 ``Sampling and reconstructing spatial fields using mobile sensors,"
 {\it Proceedings of the  IEEE International Conference on Acoustics, Speech and Signal Processing (ICASSP), 2012}, pp. 3789--3792, 25-30 March 2012.

\bibitem{UnnVet12b}J.~Unnikrishnan and M.~Vetterli,
 ``On sampling a high-dimensional bandlimited field on a union of shifted lattices," 
  {\it Proceedings of the IEEE International Symposium on Information Theory (ISIT), 2012}, pp. 1468--1472, 1-6 July 2012.

\bibitem{UnnVet12c}
J.~Unnikrishnan and M.~Vetterli,
 ``Sampling High-Dimensional Bandlimited Fields on Low-Dimensional Manifolds," {\it IEEE Trans. Inform. Theory}, vol.~59, no.~4, pp. 2103--2127, April 2013.
 

\bibitem{WeiVet12}
C.~Weidmann and M.~Vetterli,
 ``Rate distortion behavior of sparse sources," 
 {\it  IEEE Trans. Inform. Theory}, vol.~58, no.~8, pp. 4969--4992, Aug. 2012.
 
 \bibitem{WuVer12}
Y.~Wu and S.~Verd{\'u},
 ``Optimal phase transitions in compressed sensing," 
 {\it IEEE Trans. Inform. Theory}, vol.~58, no.~10, pp. 6241--6263, Oct. 2012.

 \bibitem{WuVer10}
Y.~Wu and S.~Verd{\'u},
 ``R{\'e}nyi information dimension: fundamental limits of almost lossless analog compression," 
 {\it IEEE Trans. Inform. Theory}, vol.~56, no.~8, pp. 3721--3748, Aug. 2010.
 
\bibitem{YamIto80}
H.~Yamamoto and K.~Itoh,
 ``Source coding theory for multiterminal communication systems with a remote source,'' 
 {\it IEICE Trans.}, vol. E63-E, no.~10, pp. 700--706, 1980.
 
\end{thebibliography}
\end{document}